\documentclass[journal]{IEEEtran}
\usepackage{cite}
\ifCLASSINFOpdf
  \usepackage[pdftex]{graphicx}
\else
  \usepackage[dvips]{graphicx}
\fi
\usepackage{amssymb,amsmath}
\usepackage{algorithm}
\usepackage{algorithmic}
\usepackage{tabularx}
\usepackage{amsthm,amssymb,amsfonts}
\usepackage{epsfig}
\usepackage{subfigure}
\usepackage{array}
\usepackage{changes}
\usepackage{textcomp}
\usepackage{xcolor}
\usepackage{amsmath}
\usepackage{fancyhdr}

\usepackage{dsfont}
\usepackage{amsmath,amssymb,amsfonts}
\usepackage{subfigure}
\usepackage{textcomp}
\usepackage{xcolor}
\usepackage{multirow}
\usepackage{mathtools}
\usepackage{mhchem}
\usepackage{tensor}
\usepackage{graphicx}  % Written by David Carlisle and Sebastian Rahtz
\usepackage{psfrag}    % Written by Craig Barratt, Michael C. Grant,
\usepackage{amsthm}
\usepackage{mathrsfs}
\usepackage{epstopdf}
\usepackage{bm}
\usepackage{cuted}
\usepackage{color, soul, framed}

\usepackage{xcolor}

\usepackage{xpatch}
\usepackage{tabularx}
\ifCLASSOPTIONcompsoc
  \usepackage[caption=false,font=normalsize,labelfont=sf,textfont=sf]{subfig}
\else
  \usepackage[caption=false,font=footnotesize]{subfig}
\fi
\usepackage{fixltx2e}
\usepackage{dblfloatfix}
\usepackage{multirow}
\usepackage{booktabs}
\usepackage{url}
\begin{document}

%\title{Waveform Design for Integrated SAR and Communication on HAPs Based on Frequency Index Modulation}
\title{Design of Frequency Index Modulated Waveforms for Integrated SAR and Communication on High-Altitude Platforms (HAPs)}
\author{Bang~Huang,~\IEEEmembership{Graduate Student Member,~IEEE,}
Sajid Ahmed,~\IEEEmembership{Senior Member,~IEEE,}
	\\Mohamed-Slim Alouini,~\IEEEmembership{Fellow,~IEEE,}
	
	%\thanks{{\color{red}This work was supported by . }(Corresponding author: Bang Huang)
	%}% <-this % stops a space
	\thanks{ The authors are with the Computer, Electrical and Mathematical Science and Engineering (CEMSE) division in King Abdullah University of Science and Technology (KAUST), Thuwal 6900, Makkah Province, Saudi Arabia.(Emails: bang.huang@kaust.edu.sa; sajid.ahmed@kaust.edu.sa; slim.alouini@kaust.edu.sa) (Corresponding author: Bang Huang)} }
\maketitle

\begin{abstract}
This paper, addressing the integration requirements of radar imaging and communication for High-Altitude Platform Stations (HAPs) platforms, designs a waveform based on linear frequency modulated (LFM) frequency-hopping signals that combines synthetic aperture radar (SAR) and communication functionalities. Specifically, each pulse of an LFM signal is segmented into multiple parts, forming a sequence of sub-pulses. Each sub-pulse can adopt a different carrier frequency, leading to frequency hops between sub-pulses. This design is termed frequency index modulation (FIM), enabling the embedding of communication information into different carrier frequencies for transmission. To further enhance the data transmission rate at the communication end, this paper incorporates quadrature amplitude modulation (QAM) into waveform design. 
%For the SAR portion, this approach reduces the ADC sampling requirements while maintaining range resolution.
The paper derives the ambiguity function of the proposed waveform and analyzes its Doppler and range resolution, establishing upper and lower bounds for the range resolution. In processing SAR signals, the receiver first removes QAM symbols, and to address phase discontinuities between sub-pulses, a phase compensation algorithm is proposed to achieve coherent processing. For the communication receiver, the user first performs de-chirp processing and then demodulates QAM symbols and FIM index symbols using a two-step maximum likelihood (ML) algorithm. Numerical simulations further confirm the theoretical validity of the proposed approach.
\end{abstract}


\IEEEpeerreviewmaketitle

\section{Introduction}

\IEEEPARstart{T}{he} rapid advancement of high-altitude platforms (HAPs) has unlocked new possibilities for wide-area surveillance, environmental monitoring, and communication applications \cite{belmekkiCellularNetworkSky2024,kurtVisionFrameworkHigh2021,SongLopez2024HighAltitudePlatformStations,dOliveira2016highaltitude,li2012high}. Operating in the stratosphere (approximately 20 km above the ground), HAPs platforms bridge the capabilities of satellites and terrestrial systems, offering the expansive coverage of satellites with the agility and lower latency of ground-based systems \cite{Renga2022CanHighAltitude}. These unique attributes make HAPs ideal for dual-function operations, where both synthetic aperture radar (SAR) imaging \cite{wang2014high} and communication \cite{YuZhang2024JointResourceAllocations} are essential. In scenarios such as disaster response, border surveillance, and remote sensing, integrated radar-communication systems deployed on HAPs can significantly enhance situational awareness while enabling real-time data transmission, positioning HAPs as invaluable assets for both civilian and defense applications \cite{Ahrazoglu2024MultiHAPs,Turk2024Design}.

Historically, radar and communication systems on shared platforms have been developed as distinct modules, each with unique signal structures and operational requirements. This separation, however, often leads to inefficiencies in spectrum and power utilization, as well as increased complexity in signal processing and hardware demands \cite{LiuMasouros2021ATutorial,ma2020joint}. As spectrum resources become increasingly limited, the demand for more efficient and compact systems has become critical. Consequently, achieving seamless integration of radar and communication functions within a single waveform has emerged as a major research priority \cite{bicua2019multicarrier,LiHu2024AoIAware}. This approach seeks to maximize spectral efficiency and minimize hardware redundancy, all while maintaining robust performance for both functionalities \cite{zhang2021overview,ChakravarthiAshoka2024ASurveyon}.

In recent years, hybrid waveform design has emerged as a promising solution to the challenges in integrated sensing and communication (ISAC) systems \cite{Ma2024IntegratedSensing}. By enabling both radar and communication functions within a single waveform, integrated designs enhance resource efficiency and simplify system architecture \cite{LiuLiao2017AdaptiveOFDM,XiaoZeng2022WaveformDesign}. Among various waveforms, Orthogonal Frequency Division Multiplexing (OFDM)—widely used in previous mobile communication generations—has garnered considerable research interest for ISAC applications \cite{gaudio2019performance,xiao2024novel,sturm2009ofdm,hsu2021analysis,huang2015low,wu2024low,liu2024ofdm}. 

Sturm \textit{et al.} \cite{sturm2009ofdm} conducted extensive simulations with practical parameters, demonstrating the feasibility of OFDM-based ISAC systems. Additionally, \cite{hsu2021analysis} investigated the influence of pilot power allocation and positioning on ISAC performance, offering targeted solutions. To mitigate the high peak-to-average power ratio (PAPR) inherent to OFDM signals, \cite{huang2015low} and \cite{wu2024low} introduced low-PAPR OFDM waveforms using iterative least squares and ADMM-based algorithms, respectively. Moreover, \cite{xiao2024novel,wei2023waveform,li2024mimo} explored combining OFDM with multi-input multi-output (MIMO) technology to form MIMO-OFDM ISAC systems, analyzing their performance benefits. These OFDM-based ISAC advancements are thus positioned to play a pivotal role in next-generation mobile and Wi-Fi communications.

In addition to OFDM, frequency hopping (FH) technology has gained attention for its effective interference avoidance capabilities \cite{boroujeni2024enhancing,wu2021frequency,wang2019co,hassanien2017dual}. Notably, Wang and Hassanien \cite{wang2019co,hassanien2017dual} proposed MIMO ISAC schemes utilizing frequency hopping to transmit information through frequency coding. Building on this, \cite{wu2021frequency} combined multi-antenna and frequency hopping techniques to further enhance ISAC system performance. Recently, advanced communication technologies, such as orthogonal time frequency space (OTFS) \cite{gaudio2019performance,yuan2022orthogonal} waveforms and reconfigurable intelligent surfaces (RIS) \cite{zhu2023joint,zhu2023joint}, have also attracted significant research interest for ISAC applications. However, it is essential to recognize that waveform design rooted in communication frameworks can inherently introduce interference, potentially affecting radar performance.

This naturally highlights the potential of radar-specific waveforms for developing integrated sensing and communication (ISAC) or joint radar-communication (RadCom) systems.\footnote{RadCom refers to radar technology focused exclusively on sensing, while ISAC integrates radar, vision, and other methods to facilitate sensing.} Widely utilized radar waveforms, such as linear frequency modulated (LFM) and frequency modulated continuous wave (FMCW) signals, are valued for their robustness to Doppler shifts during matched filtering, ability to deliver high range resolution, and improved signal-to-noise ratio through pulse compression. However, traditional LFM and FMCW signals lack the inherent modulation structures necessary for efficient data communication.
To address this, researchers have been exploring integrated waveform designs for ISAC that are based on LFM/FMCW signals \cite{ma2021frac,huang2020majorcom,barrenechea2007fmcw,gu2022design,xie2021waveform,ma2021spatial}. In \cite{barrenechea2007fmcw}, an approach is proposed that employs amplitude modulation (AM) with FMCW waveforms, enabling communication without compromising radar performance. Meanwhile, \cite{xie2021waveform} presents an ISAC waveform based on multiple phase-shift keying (MPSK) LFM, balancing radar and communication capabilities by evaluating key metrics such as energy leakage, peak-to-side lobe ratio (PSLR), and transmission rate. 

Index modulation (IM), a promising technology for next-generation energy-efficient communications, has also been explored for enhancing ISAC. By modulating parameters like phase and frequency within LFM or FMCW waveforms using IM, it is possible to achieve communication with minimal impact on radar performance while improving spectral utilization and transmission rates. For instance, in \cite{gu2022design}, a frequency IM (FIM)-FMCW waveform is developed, indexing distinct carrier frequencies of FMCW sub-pulses and combining them with amplitude and phase modulation (APM) techniques. Both theoretical and empirical analyses confirm that this method preserves radar detection performance while enhancing communication quality. 
Moreover, Ma \textit{et al.} \cite{ma2021frac} integrate IM with sparse arrays and narrowband waveforms, allowing high-resolution radar sensing and superior communication performance with reduced hardware complexity. Additionally, in \cite{huang2020majorcom}, an innovative ISAC system called multi-carrier agile joint radar communication (MAJoRCom) is proposed based on carrier agile phased array radar (CAESAR). This approach utilizes spatial and spectral agility to enable digital message transmission without sacrificing radar performance, achieving efficient integration of radar and communication functionalities.  

In the ISAC solutions discussed above, sensing modules primarily target detection and position estimation. However, as research advances and sensing applications expand, the ability to generate a comprehensive image of the surrounding environment while supporting communication is gaining significant industry interest \cite{liu2023integrated,manzoni2024evaluation,lahmeri2022trajectory,lahmeri2024uav}. For example, utilizing HAPs to deliver communication services in remote areas while capturing crop images enables more precise monitoring and management, further promoting the development of intelligent and precision agriculture \cite{belmekkiCellularNetworkSky2024}. 
Synthetic Aperture Radar (SAR), a widely used radar technology, is increasingly explored for integration into ISAC systems. Notably, OFDM signals have emerged as a focal point of research in SAR imaging applications \cite{zhang2014ofdm,wang2019first}. In an integrated SAR and communication (ISARAC) framework, Zhang \textit{et al.} proposed using OFDM signal cyclic prefixes (CPs) to enhance SAR imaging and communication capabilities \cite{zhang2014ofdm}. Additionally, Wang \cite{wang2019first} developed an OFDM imaging approach using space-time coding (STC) configurations, which enables range-Doppler signal decoupling and differentiation between communication signals and SAR echoes. Expanding on this, Liu \textit{et al} \cite{liu2022radar} introduced a CP-free OFDM-based ISAC system that removes cyclic prefixes to reduce energy loss and mitigate false targets, achieving high-resolution SAR imaging of ground targets while maintaining a low bit error rate (BER) in the communication subsystem. 
This approach offered detailed analysis and demonstrates superior performance in an air-to-ground scenario. Furthermore, Tan \textit{et al.} introduced an innovative ISARAC architecture that utilizes time-frequency spectrum shaping (TFSS) with short-time Fourier transform (STFT) to embed information and enhance imaging performance, offering a lightweight, low-cost, and secure solution for integrating SAR and communication. Building on this, the same authors \cite{Yang2023WaveformDesign} later proposed an ISAC waveform design leveraging DFT watermarking to optimize peak sidelobe level (PSL) for covert communication and low probability of interception (LPI) radar sensing, incorporating peak-to-average ratio (PAR) and energy constraints to ensure hardware compatibility, with the approach validated through SAR imaging and communication experiments. Although the aforementioned studies have initiated research on ISARAC, existing work lacks in-depth exploration of practical application scenarios, and the designed waveforms have limited practicality. Therefore, it is essential to further investigate ISARAC waveform design in real-world application contexts to enhance its usability and relevance.

In response to these challenges, this paper proposes a novel integrated waveform based on LFM frequency-hopping (FH-LFM) signals, designed specifically for HAPs platforms to support both SAR imaging and communication. This design, termed FIM, segments each pulse of an LFM signal into sub-pulses, each with a different carrier frequency, effectively creating frequency hops between sub-pulses. By embedding communication data within the frequency indices of these sub-pulses, this approach facilitates simultaneous radar and communication functions. Importantly, the design retains the range resolution advantages of LFM for SAR applications while alleviating the analog-to-digital converter (ADC) sampling requirements, addressing a critical issue in high-resolution radar systems.

The theoretical framework of this waveform includes a detailed derivation of its ambiguity function, allowing an in-depth analysis of its Doppler and range resolution characteristics. Upper and lower bounds for range resolution are established to confirm that SAR imaging quality is maintained. To resolve phase discontinuities between sub-pulses, a phase compensation algorithm is introduced, ensuring coherent SAR signal processing. On the communication side, a two-step maximum likelihood (ML) algorithm is employed to demodulate quadrature amplitude modulation (QAM) and frequency index modulation (FIM) symbols, supporting efficient data extraction.

Numerical simulations validate the efficacy of the proposed design, demonstrating that it meets both radar and communication performance requirements. This research presents a significant advancement in the field of HAPs-based integrated radar and communication systems, offering a robust and efficient solution that could enhance the operational capabilities of HAPs platforms in multi-functional scenarios.

The structure of this paper is organized as follows: Section \ref{sec2} presents an integrated waveform designed for the ISARAC system and provides an in-depth analysis of its ambiguity function. Section \ref{sec3} outlines the design of both the SAR and communication receivers, along with the corresponding algorithms for imaging and demodulation. Following this, numerical simulations, in Section \ref{sec4} are conducted to verify the accuracy and effectiveness of the proposed algorithms. Finally, the conclusions of this paper are summarized in Section \ref{sec5}.

\section{Proposed Signal Model of ISARAC System} \label{sec2}
\begin{figure}[t]
	\centering
	{\includegraphics[width=0.35\textwidth]{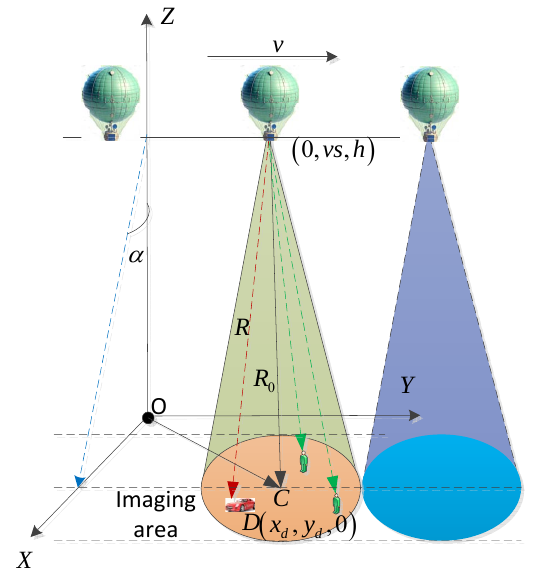}}
	\caption{Basic model of a joint communication and SAR imaging system. The HAP, at an altitude $h$ and traveling along a straight path with velocity $v$, transmits frequency-modulated sub-pulses, each with a distinct carrier frequency.}
	\label{fig1}	
\end{figure}

Consider a scenario in which a HAP at an altitude $H$ travels along a straight path with a constant velocity $V$ as shown in Fig. \ref{fig1}. In this model, we consider a strip-map \cite{JirousekPeichl2024DLR,JirousekPeichl2022SyntheticApertureRadar,JirousekPeichl2023DesignofaSynthetic} HAP that continuously transmits signals for user communication, receives reflected signals from the ground, and strategically combines these received signals to generate an image of the ground. The transmission and reception of SAR signals are both carried out by the HAP, while the transmission and reception of the communication link are handled by the HAP and the user terminal, respectively. To ensure continuous communication with users, the HAP can follow a circular path, or multiple HAPs can be deployed \cite{Ahrazoglu2024MultiHAPs}. However, this study focuses solely on a single HAP for SAR imaging at the radar end, and does not address the potential impacts of switching HAPs relays at the communication end.    

%This paper considers a scenario of integrated SAR and communication (ISARAC) imaging for HAPs, where HAPs transmits ISARAC signal to communicate with the users and meanwhile HAPs receiver will receive the reflected signal to process a SAR image. 
 % Further details can be found in Fig. \ref{fig1}\footnote{Fig. \ref{fig1} illustrates the enhancement of communication terminal performance through the deployment of multiple HAPs platforms. Furthermore, as this paper employs the stripmap imaging mode outlined in References \cite{JirousekPeichl2024DLR,JirousekPeichl2022SyntheticApertureRadar,JirousekPeichl2023DesignofaSynthetic}, the presence of these multiple HAPs platforms ensures uninterrupted communication links between users and HAPs \cite{Ahrazoglu2024MultiHAPs}. }.
% 
%This scenario may occur in areas affected by natural disasters such as earthquakes. In such cases, it is essential not only to maintain communication with affected users but also to accurately map the surrounding environment to assist relevant authorities in quickly assessing the situation and making effective response decisions. Furthermore, the remainder of this section will first detail the design process of the proposed LFM-FIM waveform, and subsequently conduct a performance analysis of its ambiguity function.
%
%
%\subsection{LFM-FIM Waveform Design for ISARAC system}
To achieve both the communication and SAR objectives, mentioned above, we divide a single linear frequency modulated (LFM) pulse into multiple segments, forming a train of sub-pulses. By assigning each sub-pulse a unique frequency band, we implement frequency hopping within the LFM signal to effectively embed communication data. Here, let $f_c$ denote the carrier frequency, $B_w$ the bandwidth, and $T_w$ the pulse width of the LFM signal. The resulting LFM waveform can then be written as:
\begin{equation}
\label{key}
s(t) = p(t)e^{j2\pi \left( f_c+\frac{1}{2}Kt \right) t},
\end{equation}
where $p(t) = \mbox{rect}\left(\frac{t}{T_w}\right)$ represents the rectangle function and $K$ is the chirp rate. 

\begin{figure}[htp]
	\centering
	{\includegraphics[width=0.35\textwidth]{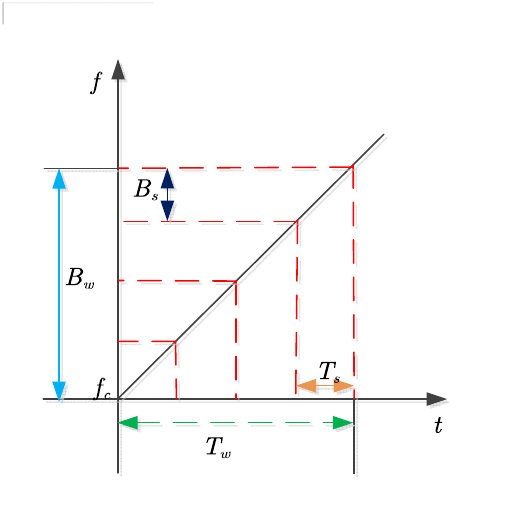}}
	\caption{Illustration for FH-LFM waveform of a chirp on time-frequency domain.}
	\label{fig2}	
\end{figure}

Further, we introduce frequency hopping to embed the communication signal into the LFM waveform. For this, LFM chirp can be divided into $M$ sub-chirps, with each sub-chirp having a bandwidth and pulse width that are \(\frac{1}{M}\) of the original chirp's bandwidth and duration, respectively, as shown in Fig. \ref{fig2}. Besides, Fig. \ref{fig2} shows that the entire transmission bandwidth is partitioned into $M$ segments, each with an equivalent bandwidth of \( B_s \). Furthermore, each sub-chirp can transmit any of these segments as needed. 

\begin{figure}[htp]
	\centering
	\subfigure[]{
			{\includegraphics[width=0.23\textwidth]{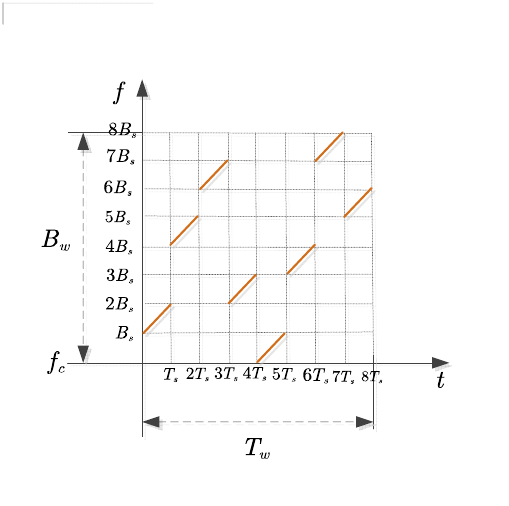}}}
	\subfigure[]{
			{\includegraphics[width=0.23\textwidth]{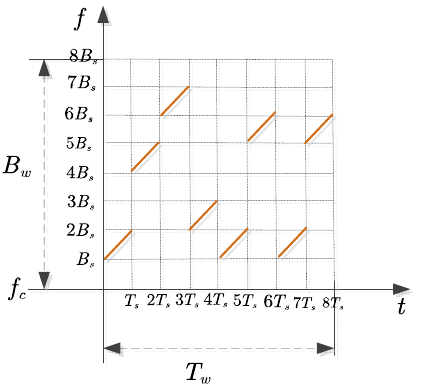}}}
	\caption{Examples of FIM-LFM signal of a chirp on time-frequency doamin with $M=8$. (a) $
		\left[ a_1,a_2,a_3,a_4,a_5,a_6,a_7,a_8 \right] =\left[ 1,4,6,2,0,3,7,5 \right] 
		$. (b) $
		\left[ a_1,a_2,a_3,a_4,a_5,a_6,a_7,a_8 \right] =\left[ 1,4,6,2,1,5,1,5 \right] 
		$.}
	\label{fig3}
\end{figure}

Denote $
B_s=\frac{B_w}{M}$, $T_s=\frac{T_w}{M}
$, the transmitted signal for $m$th sub-chirp can be expressed as
\begin{equation}
	\label{key}
	\begin{split}
	s_{m}\left( t \right) = p\left(t-\bigtriangleup t_m\right)\sqrt{P}e^{j2\pi \left( f_c+a_mB_s+\frac{1}{2}K\left( t-\bigtriangleup t_m \right) \right) \left( t-\bigtriangleup t_m \right)}.
	\end{split}
\end{equation}
with $m=0,1,...,M-1$ and $
\bigtriangleup t_m=mT_s
$. Further, ${P}$ stands for the transmit power of ISARAC signal and $
a_m\in \left\{ 0,1,2,....,M-1 \right\} 
$ is the index of the sub-bandwidth in the sub-chirp. Therefore, the modulation of the communication signal is achieved by hopping the frequency of the transmitted signal.  In fact, the frequency variation of the sub-chirp transmitted signal is accomplished by indexing different signal bandwidths. Hence, this paper designates it as frequency-index modulation (FIM). As provided in Fig. \ref{fig3}, the transmit bits for every pulse is $
p_f=M\log _2M
$.

Meanwhile, to further enhance the transmission rate of the communication signal, $J$-QAM will also be utilized. Consequently, a signal incorporating both QAM-FIM modulation and the LFM waveform can be expressed as:
\begin{multline}
\label{key}
s_{m}\left( t \right)=p\left(t-\bigtriangleup t_m\right) c_{m}\sqrt{P}\\	
\times e^{j2\pi \left( f_c+a_mB_s+\frac{1}{2}\left( t-\bigtriangleup t_m \right) \right) \left( t-\bigtriangleup t_m \right)}.
\end{multline}
with $c_{m}=A_me^{j\phi _m}$ denotes the QAM symbol for the sub-chirp. Note that this paper assumes that $A_m$ and $\phi_m$ are the amplitude and phase rotation. In this respect, each pulse can carry $p=p_f+M\log _2J$ bits of communication information.

Moreover, the expression for ISARAC signal consisting of $K$ pulses can be expressed as 
\begin{equation}
	\label{eq4}
	s\left( t \right) =\sum_{k=0}^{K-1}{\sum_{m=0}^{M-1}{s_{k,m}\left( t \right)}}
\end{equation}
with the sub-pulse  
\begin{multline}
\label{key}
s_{k,m}(t)=p\left(t-\bigtriangleup t_{kM+m}\right)c_{kM+m}\sqrt{P}\\
\times e^{j2\pi \left( f_c+a_{kM+m}B_s+\frac{1}{2}K\left( t-\bigtriangleup t_{kM+m} \right) \right) \left( t-\bigtriangleup t_{kM+m} \right)}.
\end{multline}
\subsection{Analysis of Ambiguity Function}  
The ambiguity function of the proposed FIM-LFM waveform can be expressed as:
\begin{equation}
\label{eq6}
\chi \left( \tau ,\xi \right) =\int_{-\infty}^{+\infty}{\tilde{s}\left( t \right) \tilde{s}^*\left( t+\tau \right) \mathrm{e}^{j2\pi \xi t}}dt
\end{equation}
where $\tau$ is the time delay and $\xi$ is the Doppler shift. The transmitted signal over the duration 
can be written as
\begin{equation}
\label{key7}
\tilde{s}\left(t\right)=\sum_{m=0}^{M-1}p\left(t-\bigtriangleup t_m\right)e^{j2\pi \left( f_c+a_mB_s+\frac{1}{2}K\left( t-\bigtriangleup t_m \right) \right) \left( t-\bigtriangleup t_m \right)}.
\end{equation}
It should be noted that in \eqref{key7} QAM symbols are not considered. This is because, as detailed in subsequent sections of this paper, QAM symbols are excluded in SAR imaging process. Therefore, their exclusion does not affect the performance of the proposed waveform when applied to SAR images, particularly in the range and Doppler dimensions.

%
%\eqref{eq9} can be further deduced as 
%\begin{equation}
%	\label{key}
%	\begin{split}
%			\check{\chi}\left( \tau ,\xi \right) =&b_1\sin\mathrm{c}\left[ 2\left( \xi +\left( a_m-a_{m^{\prime}} \right) B_s-K\tau \right) \right. 
%		\\
%	&	\times \left. 
%	\left( T_s-\left| \xi +\left( a_m-a_{m^{\prime}} \right) B_s-K\tau \right| \right) 
%	 \right], 
%	\end{split}
%\end{equation}
%with 
%\begin{equation}
%	\label{key}
%	\begin{split}
%		b_1=&\left[ T_s-\left| \xi +\left( a_m-a_{m^{\prime}} \right) B_s-K\tau \right| \right] \\
%		&\times e^{j\pi \left( \xi +\left( a_m-a_{m^{\prime}} \right) B_s-K\tau \right) \left( \left( \bigtriangleup t_{m^{\prime}}-\bigtriangleup t_m \right) -\tau \right)}.
%	\end{split}
%\end{equation}
%{\color{red}
%Further, as demonstrated in the appendix \ref{refA}, the magnitude of \(\chi(\tau, \xi)\) can be approximated as: 
The magnitude of the ambiguity function of $\tilde{s}(t)$ can be derived as
\begin{equation}
	\label{eq15}
	\begin{split}
		\left|{\chi}\left( \tau ,\xi \right) \right|=& \left( T_s-\left| \tau \right| \right) \operatorname{sinc}\left( \pi\left( \xi -K\tau \right) \left( T_s-\left| \tau \right| \right) \right) 
		\\
		&\times \left|\sum_{m=1}^M{e^{j2\pi \left[ \xi mT_s-a_m B_s\tau \right]}} \right|.
	\end{split}
\end{equation}
Please see Appendix \ref{refA} for the proof.
\subsubsection{Doppler Resolution}
Inserting $\tau=0$ in \eqref{eq15} yields the Doppler profile of the ambiguity function as
\begin{equation}
	\label{0}
	\left| \tilde{\chi}\left( 0,\xi \right) \right|=\left| T_s\operatorname{sinc}\left( \pi \xi T_s \right) \frac{\sin \left( \pi M\xi T_s \right)}{\sin \left( \pi \xi T_s \right)} \right|.
\end{equation} 
Thus, the Doppler resolution of the proposed FIM-LFM waveform is 
\begin{equation}
	\label{eq17}
	\rho _{\xi}=\frac{1}{MT_s}=\frac{1}{T_w}, 
\end{equation}
which is consistent with that of the original LFM signal.
\subsubsection{Range profile}
Similarly, inserting $\xi=0$ in \eqref{eq15} results in the range profile of ambiguity function  
\begin{multline}
\left|\tilde{\chi}\left(\tau,0\right)\right| =\bigg|\left( T_s-\left| \tau \right| \right) \operatorname{sinc}\left( \pi K\tau \left( T_s-\left| \tau \right| \right)\right) \\
			 \times \sum_{m=1}^M{e^{-j2\pi a_m B_s\tau}} \bigg|. \label{eq18}
\end{multline}
It should be noted that $a_m$ can change randomly, which makes it challenging to find the range profile. However, we can put a lower and upper bound on the range resolution.
\begin{figure}[htp]
	\centering
	\subfigure[]{
		{\includegraphics[width=0.23\textwidth]{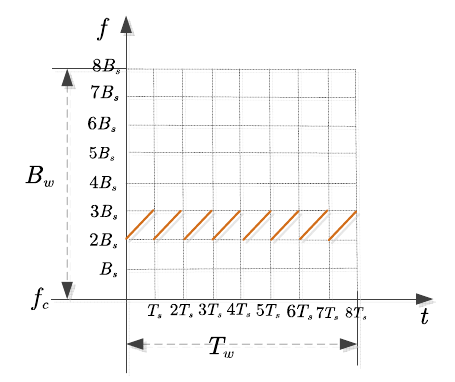}}}
	\subfigure[]{
		{\includegraphics[width=0.23\textwidth]{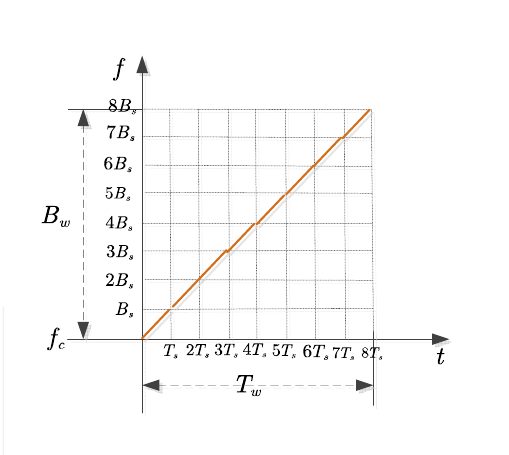}}}
	\caption{Examples of FIM-LFM signal of a chirp on time-frequency doamin with $M=8$. (a) $
		a_m=2,m=1,2,...,M
		$
		. (b) $
		a_m=m,m=1,2,...,M
		$.}
	\label{fig4}
\end{figure}

If all sub-pluses choose the same band, as shown in Fig.\ref{fig4}(a), it will correspond to the minimum bandwidth of the overall signal. In this case, (\ref{eq18}) can be re-written as 
\begin{equation}
	\label{eq19}
	\left| \tilde{\chi}\left( \tau ,0 \right) \right|=\left| \left( T_s-\left| \tau \right| \right) \operatorname{sinc}\left( \pi K\tau \left( T_s-\left| \tau \right| \right) \right)  \right|.
\end{equation}
The range ambiguity function expressed in \eqref{eq19} matches that of an LFM signal with a pulse width $T_s$ and bandwidth $B_s$ and it can be readily derived as
%It is obvious from \eqref{eq19} that the range resolution capability for proposed waveform is 
\begin{equation}
	\label{key}
	\rho _r=\frac{c}{2B_s},
\end{equation}
which is an upper bound on the range resolution.  
%Compared to the original LFM signal, the range resolution of the proposed waveform is halved under this condition.
%}

Similarly, for the case when $a_m=m$ shown in Fig.\ref{fig4}(b), the spectra of adjacent sub-pulses are continuous. This case will correspond to the maximum bandwidth of the overall signal.  Now, the range profile of the ambiguity function can be expressed as
\begin{equation}
\label{eq21}
\left|\tilde{\chi}\left( \tau ,0 \right) \right|=\left|\left( T_s-\left| \tau \right| \right) \operatorname{sinc}\left( \pi K\tau \left( T_s-\left| \tau \right| \right) \right) \frac{\sin \left( \pi MB_s\tau \right)}{\sin \left( \pi B_s\tau \right)} \right|.
\end{equation}
Using \eqref{eq21} the range resolution of the FIM-LFM signal can be derived as
\begin{equation}
\label{key}
\rho _r=\frac{c}{2MB_s}=\frac{c}{2B_w},
\end{equation}
which is the minimum bound on the ambiguity function. In this case, the range resolution of the proposed waveform is consistent with that of the LFM waveform with the pulse width and bandwidth being \( T_s \) and \( B_s \), respectively. Therefore, we can say that the range resolution  
\begin{equation}
	\label{eq16}
	\frac{c}{2B_s}\leqslant \rho _r\leqslant \frac{c}{2B_w}.
\end{equation}
This implies that, due to the frequency hopping of adjacent sub-pulses, the range resolution is no longer a fixed value.
\section{Receiver design for ISARAC} \label{sec3}
In this section, we will analyze the SAR and communication receivers of the ISARAC system.

\subsection{SAR receiver}
Using \eqref{fig1} the received signal by the SAR can be written as 
\begin{equation}
\label{key}
r\left( t \right) =\sum_{k=0}^{K-1}{\sum_{m=1}^M{s_{k,m}\left( t,\tau \left( s \right) \right)}},
\end{equation}
where 
\begin{eqnarray}
\label{key}
s_{k,m}\left( t,\tau \left( s \right) \right) &=& p\left(t-\bigtriangleup t_{kM+m}-\tau(s)\right) \sigma c_{kM+m} \notag\\
&&\times~e^{j2\pi \left[ f_c+a_{kM+m}B_s \right] \left( t-\bigtriangleup t_{kM+m}-\tau \left( s \right) \right)}\notag\\
&&~~\times~e^{j\pi K\left( t-\bigtriangleup t_{kM+m}-\tau \left( s \right) \right) ^2}.
\end{eqnarray}
while $\sigma$ denotes the radar cross section (RCS) and $\tau(s)$ represents the time delay experienced by the signal as it travels from the transmitter to the target and then reflects to the SAR receiver. The time delay $\tau(s)=\frac{2R\left( s \right)}{c}$ with $R(s)$ representing the range between HAP and the ground scatterer at slow time $s$ can be derived using Fig. \ref{fig1} as 
\begin{equation}
	\label{eq26}
	\begin{split}
			R\left( s \right) =&
			\sqrt{x_{d}^{2}+\left( y_d-vs \right) ^2+h^2}
			\\
			\approx & 
			R_{0}^{} + \frac{\left( y_d-vs \right)^2}{2R_{0}}			,
	\end{split}
\end{equation}
where $R_{0} = \sqrt{x_{d}^{2}+h^2}$ is the closest distance between the HAPs and the center of the imaging scene and $v$ denotes the flight speed of the HAPs. Besides, The 3D Cartesian coordinates of the target and the HAPs are represented as \((x_d, y_d, 0)\) and \((0, vs, h)\), respectively. It should be noted that the second approximation in Eq.\eqref{eq26} is obtained through a second-order Taylor expansion.

%Define $
%\tau _0=\frac{2R_{0}^{}}{c}$, $\bigtriangleup \tau\left( s \right) =\frac{\left( Vs \right) ^2}{cR_{0}^{}}
%$. Then we have 
%\begin{equation}
%	\label{key}
%	\tau \left( s \right) =\tau _0+\bigtriangleup \tau \left( s \right) 
%\end{equation}

The received sub-pulse at the SAR receiver after demodulating by the carrier frequency $f_c$ and removing the QAM symbols\footnote{QAM symbols transmitted by the SAR are known at the SAR receiver.} can be written as
%After the aforementioned operation and then removing the QAM symbol, the received signal of the sub-pulse becomes: 
%
%{\color{red}
\begin{equation}
\label{eq28}
\begin{split}
\tilde{s}_{k,m}(t,\tau (s))=&p(t-\bigtriangleup t_{kM+m}-\tau (s))\sigma e^{-j2\pi f_c\tau \left( s \right)}
\\
&\times e^{j2\pi a_{kM+m}B_s\left( t-\bigtriangleup t_{kM+m}-\tau \left( s \right) \right)}
\\
&\times e^{j\pi K\left( t-\bigtriangleup t_{kM+m}-\tau \left( s \right) \right) ^2}.  
\end{split}
\end{equation}
%}
%
The phase of the third term will be much less than the first two terms. Therefore, ignoring the phase of the third term, the Doppler frequency of the signal in \eqref{eq28} can be easily obtain:
\begin{multline}
\label{eq29}
f(s) = \frac{\left( f_c+a_{kM+m}B_s\right)(y_d-vs)v}{cR_0}\\
     = \frac{\left( f_c+a_{kM+m}B_s\right)y_dv}{cR_0} - \frac{\left( f_c+a_{kM+m}B_s\right)v^2s}{cR_0}.
\end{multline}
The presence of the second term in \eqref{eq29} results in a varying signal frequency within a single pulse. Without proper compensation, directly performing azimuth compression is highly likely to cause defocusing of the azimuth signal, preventing the generation of a clear SAR image.

Applying an FFT transform to \eqref{eq28} along the range dimension yields
\begin{multline}
\label{eq299}
S_{k,m}\left( f_t,\tau \left( s \right) \right) = \int_{-\infty}^{\infty}{\tilde{s}_{k,m}\left( t,\tau(s)\right)}e^{-j2\pi f_tt} dt \notag\\
\quad\quad\quad\quad\quad\quad=\sigma~p\left[ \frac{f_t}{B_s} \right] \mathrm{e}^{-j\pi \frac{f_{t}^{2}}{K}}e^{-j2\pi \left( f_c+a_{kM+m}B_s \right) \tau \left( s \right)}\notag\\
\times e^{-j2\pi f_{t}^{}\left( \bigtriangleup t_{kM+m}+\tau \left( s \right) \right)}
\end{multline}
where $f_{t}$ denoted the range frequency.
Furthermore, the following correction factors are introduced for performing pulse compression and time-domain alignment:
\begin{equation}
	\label{eq30}
	H_{kM+m}^{rc}=p\left[ \frac{f_t}{B_s} \right] \mathrm{e}^{-j\pi \frac{f_{t}^{2}}{K}}e^{j2\pi f_{t}^{}\bigtriangleup t_{kM+m}}
\end{equation}

Multiplying the sub-band echo spectrum by \eqref{eq30} to obtain the sub-band spectrum after pulse compression, given as
\begin{equation}
	\label{eq31}
	\begin{split}
			S_{k,m}^{rc}\left( f_t,\tau \left( s \right) \right) =&S_{k,m}\left( f_t,\tau \left( s \right) \right) H_{kM+m}^{rc}
		\\
		=&\sigma p\left[ \frac{f_t}{B_s} \right] e^{-j2\pi \left( f_c+a_{kM+m}B_s+f_{t}^{} \right) \tau \left( s \right)}
	\end{split}
\end{equation}

From \eqref{eq31}, it can be seen that each sub-band's frequency is shifted by \( a_{kM+m}B_s \). Therefore, to maintain coherence in the azimuth dimension, a shift of \( a_{kM+m}B_s \) can be applied in the range frequency dimension. After performing the aforementioned operations, the following is obtained:
\begin{equation}
	\label{key}
	\begin{split}
		&S_{k,m}^{rc}\left( f_t-a_{kM+m}B_s,\tau \left( s \right) \right) =
		\\
		&\qquad\qquad\quad\sigma p\left[ \frac{f_t-a_{kM+m}B_s}{B_s} \right] e^{-j2\pi \left( f_c+f_{t}^{} \right) \tau \left( s \right)}
	\end{split}
\end{equation}

Next, summing all the sub-pulses of the \( k \)th pulse results in
\begin{equation}
	\label{key}
	\begin{split}
		S_{\mathrm{sum}}=&\sum_{m=1}^M{S_{k,m}^{rc}\left( f_t-a_{kM+m}B_s,\tau \left( s \right) \right)}
		\\
		=&\sigma e^{-j2\pi \left( f_c+f_{t}^{} \right) \tau \left( s \right)}\sum_{m=1}^M{p\left[ \frac{f_t-a_{kM+m}B_s}{B_s} \right]}
	\end{split}
\end{equation}

Next, by transforming the bandwidth-synthesized signal from the range frequency domain into the fast time domain, the following result is obtained:
\begin{eqnarray}
s(t,\tau(s))\!\!\!\!&=&\!\!\!\!\int_{-\infty}^{\infty}S_{\mathrm{sum}}e^{j2\pi f_tt }df_t \notag\\
		    \!\!\!\!&=&\!\!\!\!\sigma e^{-j2\pi f_c\tau(s)}\sum_{m=1}^M\int_{-\infty}^{\infty}e^{j2\pi f_t(t-\tau(s))} \notag\\
		  \!\!\!\!&&\!\!\!\!\times p\left[ \frac{f_t-a_{kM+m}B_s}{B_s} \right] df_t  \notag \\
		  \!\!\!\!\!\!\!\!&=&\!\!\!\!\sigma e^{-j2\pi f_c\tau(s)}\!\sum_{m=1}^M\!{\int_{-\frac{B_s}{2}+a_{kM+m}B_s}^{\frac{B_s}{2}+a_{kM+m}B_s}\!\!{e^{j2\pi f_{t}^{}\left( t-\tau(s)\right)}}}df_t \notag \\
		  \!\!\!\!&=&\!\!\!\!\sigma B_se^{-j2\pi f_c\tau \left( s \right)}\operatorname{sinc}\left[ \pi B_s\left( t-\tau \left( s \right) \right) \right] \notag\\
		&&\!\!\!\!\times \sum_{m=1}^M{e^{j2\pi a_{kM+m}B_s\left( t-\tau \left( s \right) \right)}} \label{eq34}
\end{eqnarray}
with $\operatorname{sinc}(x) =\frac{\sin x}{x}$ representing the sinc function.
%
%For the received signal with a time shift $-
%a_{kM+m}T_s
%$, its phase can be expressed as:
%\begin{equation}
%	\label{eq30}
%	\begin{split}
%		\varphi \left( s \right) =&-2\pi \left( f_c+a_{kM+m}B_s \right) \tau \left( s \right) 
%		\\
%		&\qquad\qquad+\pi K\left( t-a_{kM+m}T_s-\tau \left( s \right) \right) ^2
%		\\
%		=&-2\pi \left( f_c+a_{kM+m}B_s \right) \tau \left( s \right) 
%		\\
%		&\qquad+\pi K\left( t-\tau \left( s \right) \right) ^2+\pi K\left( a_{kM+m}T_s \right) ^2
%		\\
%		&\qquad-2\pi Ka_{kM+m}T_st+2\pi KT_s\tau \left( s \right) a_{kM+m}
%		\\
%		=&-2\pi f_c\tau \left( s \right) +\pi K\left( t-\tau \left( s \right) \right) ^2
%		\\
%		&\qquad+\pi K\left( a_{kM+m}T_s \right) ^2-2\pi a_{kM+m}B_st
%	\end{split}
%\end{equation}
%Note that the transition from the second equation to the third equation involves the application of the conclusion derived from $
%K=\frac{B_s}{T_s}
%$.
%
%From observing \eqref{eq30}, when the third and fourth phase terms are compensated, the Doppler frequencies of all sub-pulses within a single pulse will be effectively aligned. 

Hence, this paper proposes the Algorithm \ref{algorithm1} to mitigate the impact of intra-pulse chirp on the Doppler frequencies of sub-pulses.
\begin{algorithm}
	\caption{Compensation algorithm for echo signal of sub-pulse}
	\label{algorithm1}
	\begin{algorithmic}[1]
		\STATE Input the received singal $\tilde{s}_{k,m}\left( t,\tau \left( s \right) \right)$ and $a_{kM+m}$;  
%		\FOR {$k = 0$ to $K-1$}
		\FOR {$m=1$ to $M$}
		\STATE Step 1: Performing FFT to $\tilde{s}_{k,m}\left( t,\tau \left( s \right) \right)$ along range domain yields $S_{k,m}\left( f_t,\tau \left( s \right) \right)$;
		\STATE Step 2: Multiplying the $S_{k,m}\left( f_t,\tau \left( s \right) \right)$ by \eqref{eq30} to obtain $S_{k,m}^{rc}\left( f_t,\tau \left( s \right) \right)$ ;
		\STATE Step 3: Shifting the sub-band waveforms frequency \( a_{kM+m}B_s \) yields $S_{k,m}^{rc}\left( f_t-a_{kM+m}B_s,\tau \left( s \right) \right)$;		
		\ENDFOR
%		\ENDFOR
		\STATE Step 4: By summing all the sub-pulse signals of the $k$ pulses and then performing an IFFT transform gives $s\left( t,\tau \left( s \right) \right)$;
		\STATE Output the compensated echo signal.
		%$
%		\tilde{s}_{k,m}\left( t-\tau \left( s \right) -
%		a_{kM+m}T_s \right)e^{j2\pi a_{kM+m}B_st} e^{-j\pi K\left( a_{kM+m}T_s \right) ^2}
%		$ 
	\end{algorithmic}
\end{algorithm}

%, the proposed method involves the following steps:
%\begin{itemize}
%	\item Step 1: Apply a time shift of $-
%	a_{kM+m}T_s
%	$ to the sub-pulse baseband echo $
%	\tilde{s}_{k,m}\left( t-\tau \left( s \right) \right) 
%	$, resulting in the signal $
%	\tilde{s}_{k,m}\left( t-\tau \left( s \right) -
%	a_{kM+m}T_s \right) 
%	$;
%	\item Step 2: Perform a frequency shift by $
%	e^{j2\pi a_{kM+m}B_st}
%	$ on the time-shifted signal, aligning the center frequency of the synthesized sub-pulse spectrum to \( f_c \);
%	\item Step 3: Apply phase compensation $
%	e^{-j\pi K\left( a_{kM+m}T_s \right) ^2}
%	$
%	 to the signal $
%	\tilde{s}_{k,m}\left( t-\tau \left( s \right) -
%	a_{kM+m}T_s \right)e^{j2\pi a_{kM+m}B_st} 
%	$.
%\end{itemize}

Finally, we can directly employ the Range-Doppler algorithm (RDA) like the operations of range cell migration correction (RCMC) and pulse compression in azimuth dimension to generate SAR images. 
\subsection{Communication receiver}
The waveform proposed in this paper serves a dual purpose, functioning both as a tool for SAR sensing and as a carrier for communication signals. In this context, we consider an imaging scenario that includes a communication user, assuming the transmitted communication signal propagates through a Line-of-Sight (LOS) channel. Consequently, the received signal at the communication user corresponding to the $m$th sub-pulse of the $k$th pulse can be written as:

\begin{multline}
	\label{key}
	\tilde{s}_{k,m}(t,\tau(s)) =p\left(t-\bigtriangleup t_{kM+m}-\tau(s)\right) h_{kM+m}c_{kM+m} \\
		               \quad\quad\quad\quad\quad\quad \times \sqrt{P}e^{j2\pi \left( f_c+a_{kM+m}B_s \right) \left( t-\bigtriangleup t_{kM+m}-\tau \left( s \right) \right)}
		\\
		                \times e^{j\pi K\left( t-\bigtriangleup t_{kM+m}-\tau \left( s \right) \right) ^2}+n\left( t \right), 
\end{multline}
where $h_{kM+m}$ denotes the Rayleigh fading channel impulse response, namely $
h_{kM+m}\sim {\cal N}\left(0,\sigma ^2 \right)$ and $n(t)\sim {\cal N}\left(0,N_o\right)$ represents the noise at the receiver. 
Assume that the communication user has the perfect knowledge of the channel and the frame synchronization.  If the reference signal known to the user is given by
\begin{equation}
	\label{eq36}
    \begin{split}
s_{\mathrm{rf}}\left( t,\tau _{\mathrm{ref}} \right) =&p(t-\bigtriangleup t_{kM+m}-\tau _{\mathrm{ref}})
\\
\,\,   & \times e^{j2\pi \left( f_c+\frac{1}{2}K\left( t-\bigtriangleup t_{kM+m}-\tau _{\mathrm{ref}} \right) \right) \left( t-\bigtriangleup t_{kM+m}-\tau _{\mathrm{ref}} \right)},
    \end{split}
\end{equation}
where $
\tau _{\mathrm{ref}}=\frac{R_{\mathrm{ref}}}{c}
$ with $R_{\mathrm{ref}}$
denoting the closest range between the central point target and the HAPs. The de-chirped signal received for a sub-chirp is expressed as:
\begin{equation}
	\label{key}
	\begin{split}
     \tilde{s}_{k,m}\left( t \right) =&\tilde{s}_{k,m}\left( t,\tau \left( s \right) \right) \left[ s_{\mathrm{rf}}\left( t,\tau _{\mathrm{ref}} \right) \right] ^*
		\\
		=&p\left(t-\bigtriangleup t_{kM+m}-\tau_{ref}\right) h_{kM+m}c_{kM+m}
		\\
		\,\,                  &\times \sqrt{P}e^{j2\pi a_{kM+m}B_s\left( t-\bigtriangleup t_{kM+m} \right)}+\tilde{n}\left( t \right) 
	\end{split}
\end{equation}
% {\color{red}
% \begin{eqnarray}
%     \tilde{s}_{k,m}\left( t \right) &=& h_{kM+m}c_{kM+m}\sqrt{P} \notag \\
%     && \times e^{-j2\pi \left(\left(f_c+a_{kM+m}B_s\right)\tau(s) + K\left(t-\bigtriangleup t_{kM+m}\right)\tau(s) - \frac{K\tau^2(s)}{2} \right)}\notag
% \end{eqnarray}

% }

It should be noted that the terms related to the delay \( \tau \left( s \right) \) have all been absorbed into \(h_{kM+m} \). Besides, $\tilde{n}\left( t \right)$ denotes the noise term.

To demodulate the FIM symbol, we can define a matching filter function as follows:
\begin{equation}
	\label{key}
	h\left( a_l \right) =e^{j2\pi a_lB_s\left( t-\bigtriangleup t_{kM+m} \right)}
\end{equation}

Subsequently, by correlating the de-chirped signal with multiple matching signals, we derive the results \eqref{eq39}, as shown in the top of next page. Furthermore, with the results of \eqref{eq39}, the demodulation of $a_{kM+m}$ can be expressed as 
\begin{figure*}[htp]
	\begin{equation}
		\label{eq39}
		\begin{split}
			\tilde{s}_{k,m}^{\mathrm{diff}}\left( t, a_l\right) =&\tilde{s}_{k,m}\left( t \right) \left[ h\left( a_l \right) \right] ^*
			\\
			=&\left\{ \begin{array}{l}
				p\left( \frac{t-\bigtriangleup t_{kM+m}}{T_s} \right) h_{kM+m}c_{kM+m}\sqrt{P}+\tilde{n}\left( t \right) ,a_l=a_{kM+m}\\
				p\left( \frac{t-\bigtriangleup t_{kM+m}}{T_s} \right) h_{kM+m}c_{kM+m}\sqrt{P}e^{j2\pi \left( a_{kM+m}-a_l \right) B_s\left( t-\bigtriangleup t_{kM+m} \right)},a_l\ne a_{kM+m}\\
			\end{array} \right. 
		\end{split}	
	\end{equation}
\end{figure*}
\begin{equation}
\label{key}
\hat{a}_{kM+m}=\underset{a_l=0,1,...,M-1}{arg\,\,\max}\left| \frac{1}{T_s}\int_{-\infty}^{\infty}{\tilde{s}_{k,m}^{\mathrm{diff}}\left( t,a_l \right)}dt \right|^2
\end{equation}

By utilizing the detected FIM symbol \( \hat{a}_{kM+m} \), QAM symbol detection can be performed based on the maximum likelihood (ML) approach as follows:
\begin{equation}
\label{key}
\hat{c}_{kM+m} = \underset{c_l\in\mathscr{T}}{\arg\,\,\min}\left| \tilde{s}_{k,m}^{\mathrm{diff}}\left( t,\hat{a}_{kM+m} \right) -h_{kM+m}c_l\sqrt{P} \right|^2
\end{equation}

%It should be noted that if the detected FIM symbol is correct, then the QAM symbol detection is typically accurate. However, if the detected FIM symbol is incorrect, the samples will only contain some erroneous terms accompanied by random noise, leading to incorrect QAM symbol detection. This indicates that there is an error propagation effect in the proposed sequential detection method. To avoid the issue of error propagation, the optimal approach is to jointly detect the FIM and QAM symbols within a sub-chirp. The joint maximum likelihood detection can be expressed as:

%\begin{table*}[htp]
%	\caption{The Comparisons for Date Rate Between Different Systems (Bits  Per Same Symbol Duration)}
%	\vspace{0pt}
%	\centering
%	\begin{tabular}{|c|c|c|c|c|c|c|c|c|c|}
%		\hline
%		\multicolumn{5}{|c|}{Parameters} & \multicolumn{5}{c|}{Systems} \\
%		\hline
%		$N_T$ & $N$ & $M$ & $L$& $J$ &GCIM-FORMASM & FOPIM & GSCIM & GCIM-SM & SM\\
%		\hline
%		4 & 2 & 8 & 8& 8 &	25 & 22 & 16 & 11 &5 \\
%		\hline
%		6 & 3 & 6 & 16& 8 &	43 & 27 & 31 & 13 &5 \\
%		\hline
%		8 & 4 & 8 & 16& 4 &	56 & 31 & 34 & 13 &5 \\
%		\hline
%	5 & 2 & 12 & 4& 4 &	22 & 25 & 11 & 8 &4\\
%		\hline
%	\end{tabular} 
%	\label{table2}
%\end{table*}
%
%
\subsection{Analysis for specific cases}
As shown in Eq.\eqref{eq34}, the response of the proposed waveform in the fast-time range dimension depends on the sinc function and the summation term related to \(a_{kM+m}\). This section will attempt to analyze the impact of two specific values of \(a_{kM+m}\), as shown in Fig.\ref{fig4}, on both the SAR and communication systems.
%{\color{red}
\subsubsection{$a_m=\mathrm{cont}.	$} where the symbol $\mathrm{cont}$ represesnts a constant, as shown in Fig.\ref{fig4}(a). Under these circumstances, Eq.\eqref{eq34} can be reformulated as:
\begin{equation}
	\label{eq42}
	\begin{split}
		s\left( t,\tau \left( s \right) \right) =&\sigma B_sMe^{-j2\pi \left[ f_c\tau \left( s \right) -a_{kM+m}B_s\left( t-\tau \left( s \right) \right) \right]}
		\\
		\,\,              &\times \operatorname{sinc}\left[ \pi B_s\left( t-\tau \left( s \right) \right) \right] 
	\end{split}
\end{equation}

At this point, the response of the waveform proposed in this paper is determined solely by the sinc function in terms of fast-time resolution. Further, the range resolution can be expressed as 
\begin{equation}
	\label{key}
	\rho _r=\frac{c}{2B_s}.
\end{equation}
%}

Conversely, the transmission of signals with identical frequencies leads to a complete absence of information at the communication end. This occurs because periodic signals inherently lack the capacity to convey any meaningful data.

\subsubsection{$a_m=m$} This means that the frequency variation between sub-pulses is continuous. Inserting above result into Eq.\eqref{eq34} yields 

\begin{equation}
	\label{eq43}
	\begin{split}
		s\left( t,\tau \left( s \right) \right) \,=&\sigma B_se^{-j2\pi f_c\tau \left( s \right)}\operatorname{sinc}\left[ \pi B_s\left( t-\tau \left( s \right) \right) \right] 
		\\
		\,\,  & \times \sum_{m=1}^M{e^{j2\pi mB_s\left( t-\tau \left( s \right) \right)}}
		\\
		=&\sigma B_se^{-j2\pi f_c\tau \left( s \right)}\operatorname{sinc}\left[ \pi B_s\left( t-\tau \left( s \right) \right) \right] 
		\\
		\,\,   &\times \frac{\sin \left( \pi MB_s\left( t-\tau \left( s \right) \right) \right)}{\sin \left( \pi B_s\left( t-\tau \left( s \right) \right) \right)}\\
		=&\sigma MB_se^{-j2\pi f_c\tau \left( s \right)}\operatorname{sinc}\left[ \pi MB_s\left( t-\tau \left( s \right) \right) \right] 
	\end{split}
\end{equation}

Therefore, \eqref{eq43} means that the sent signal can be regarded as equivalent to a LFM waveform with a bandwidth of \(M B_s\).

Similarly, the highly regular nature of the transmitted waveform results in a significant limitation of the valid information it can carry. From a communications standpoint, such scenarios should be avoided in practical applications. On the other hand, from a SAR perspective, this type of waveform can greatly enhance range resolution. Therefore, in practical implementations, it is crucial to strike a balance between optimizing both SAR and communication functionalities.
%Taking the modulus of \eqref{eq43} while disregarding the constant term yields the following expression:
%\begin{equation}
%	\label{key}
%	\left| s\left( t,\tau \left( s \right) \right) \right|=\left| \mathrm{A}_1\left( t,\tau \left( s \right) \right) \mathrm{A}_2\left( t,\tau \left( s \right) \right) \right|
%\end{equation}
%with $
%\mathrm{A}_1\left( t,\tau \left( s \right) \right) =\sin c\left[ \pi B_s\left( t-\tau \left( s \right) \right) \right] 
%$ and $
%\mathrm{A}_2\left( t,\tau \left( s \right) \right) =\frac{\sin \left( \pi MB_s\left( t-\tau \left( s \right) \right) \right)}{\sin \left( \pi B_s\left( t-\tau \left( s \right) \right) \right)}
%$. The symbol $
%\left| \cdot \right|
%$ denotes the modulus operator.

\section{Simulation Results}
\label{sec4}
To validate the effectiveness of the proposed waveform and corresponding algorithm, this section provides numerical simulation experiments for both SAR and communication systems independently. To elaborate, we assume the carrier frequency of the transmitted signal is 3.2 GHz, with a pulse width of 40 microseconds and a signal bandwidth of 80 MHz. Thus, the range resolution can be expressed as: $
\frac{c}{2B}=1.8750m
$. The HAPs  has an antenna size of 2 meters in the azimuth direction, operates at a flight altitude of 20 km, and travels at a speed of 100 m/s. The depression angle with respect to the imaging scene is set at 60°, and the dimensions of the imaging scene are 1000 m by 300 m. In the absence of special instructions, $M$ should be set to 4.
\begin{figure}[htp]
	\centering
			{\includegraphics[width=0.45\textwidth]{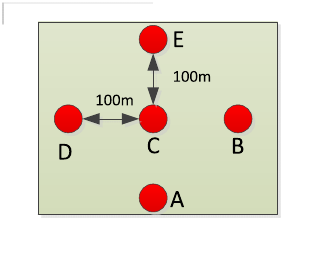}}
	\caption{SAR image depicting a scenario with five distinct point targets.}
	\label{fig5}
\end{figure}

\subsection{SAR end}
As illustrated in Fig. \ref{fig5}, suppose the SAR imaging scene contains five point targets, labeled A, B, C, D, and E. Notably, target C is positioned at the center of the scene, serving as a reference point for the layout.

\begin{figure}[htp]
	\centering
	{\includegraphics[width=0.45\textwidth]{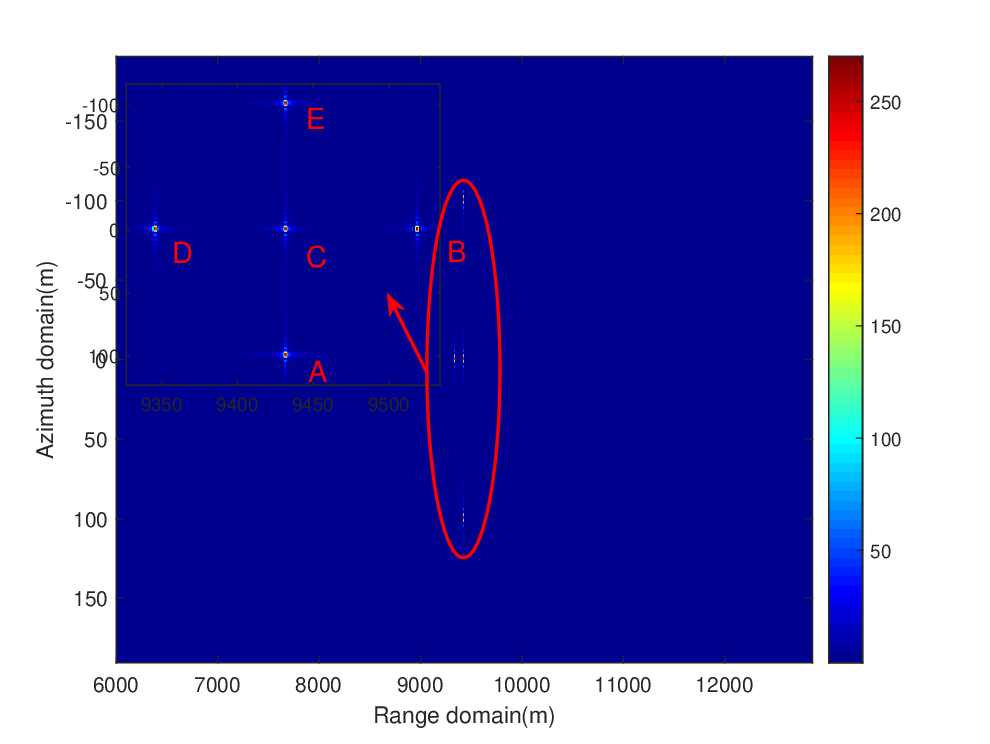}}
	\caption{SAR imaging results obtained using the LFM waveform.}
	\label{fig6}
\end{figure}

\begin{figure}[htp]
	\centering
	\subfigure[No compensation]{
		{\includegraphics[width=0.38\textwidth]{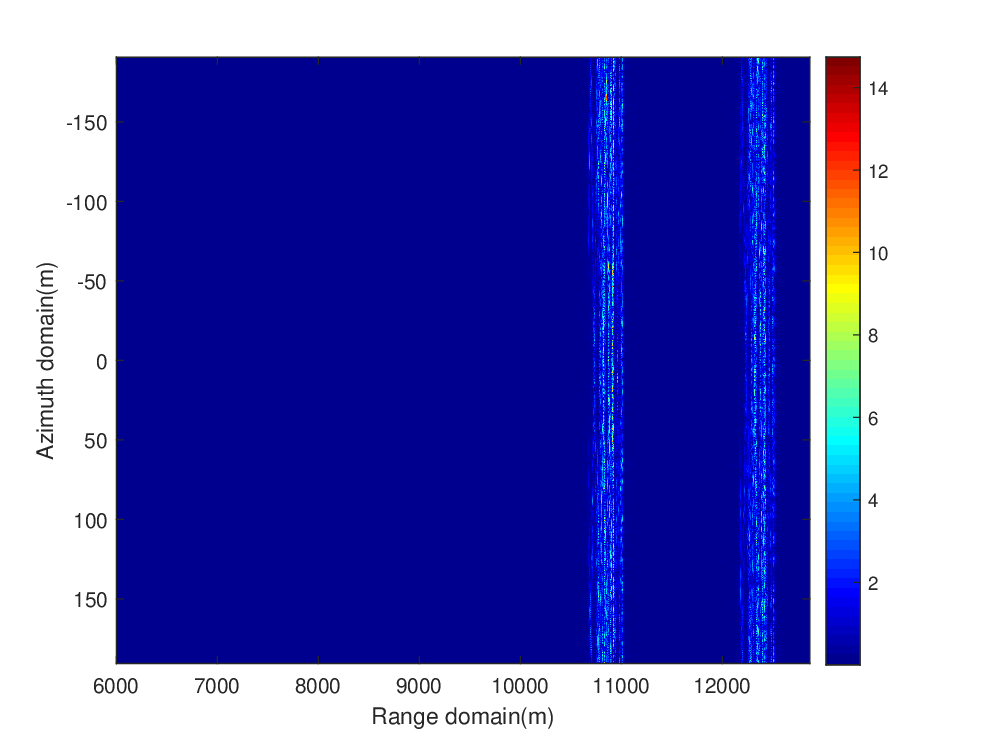}}}
	\subfigure[After removing QAM symbol]{
		{\includegraphics[width=0.38\textwidth]{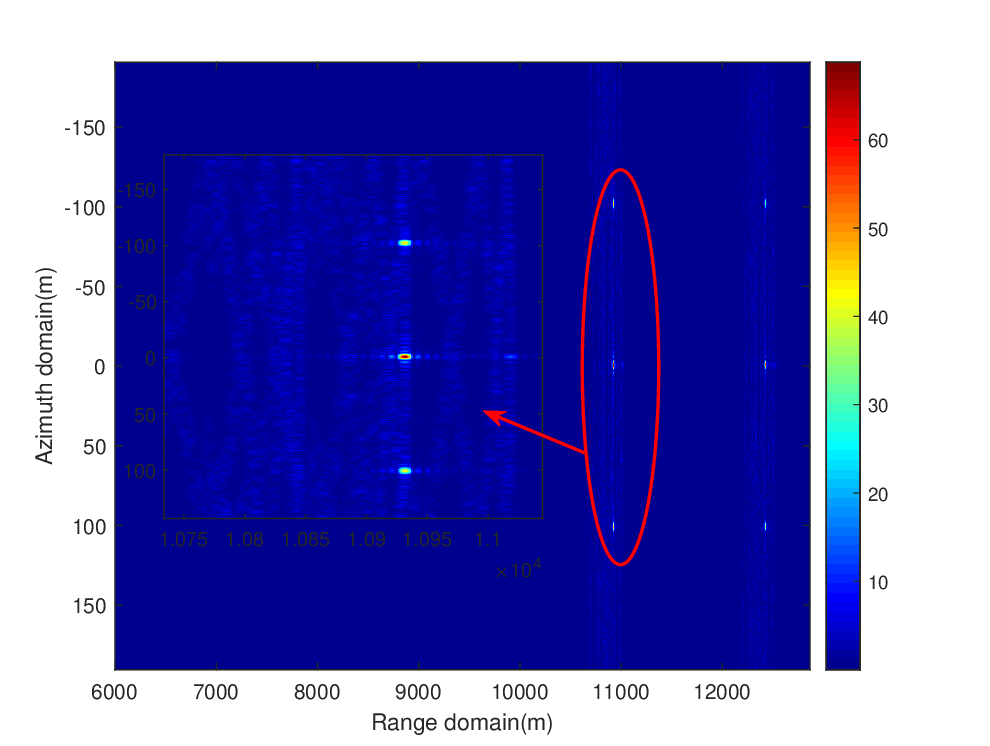}}}
	\subfigure[After applying compensation algorithm 1]{
		{\includegraphics[width=0.38\textwidth]{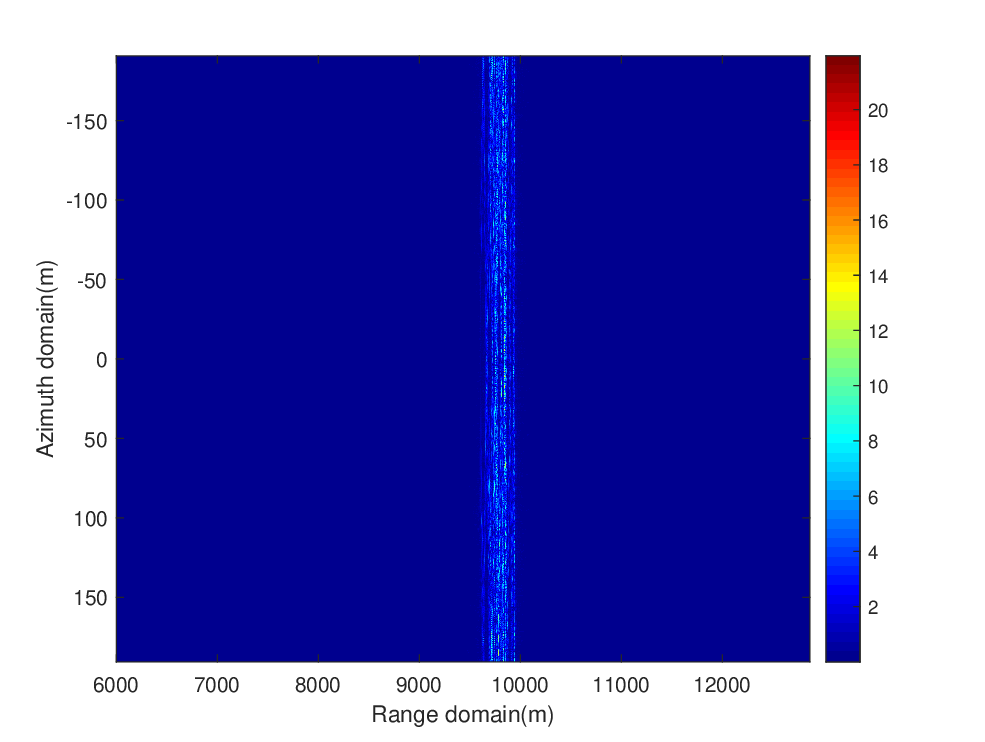}}}
	\subfigure[After removing QAM symbol and applying compensation algorithm 1]{
		{\includegraphics[width=0.38\textwidth]{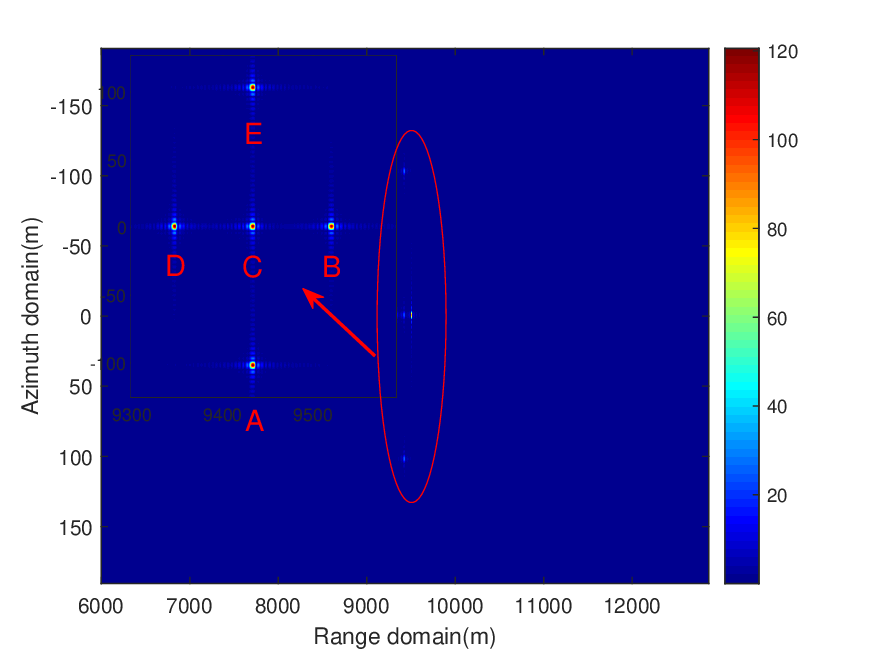}}}	
	\caption{SAR image results by applying our proposed waveform.}
	\label{fig7}
\end{figure}

Figures \ref{fig6} and \ref{fig7} show SAR images generated using the LFM waveform and the integrated FIM-LFM waveform proposed in this paper, respectively. Fig.\ref{fig7}(a) presents the result obtained by directly applying the RD algorithm without removing the QAM symbols or performing Doppler compensation. It can be seen that the SAR image generated by the unprocessed FIM-LFM waveform is severely defocused, and even two strips appear. Further analysis is shown in Fig.\ref{fig7}(b), where the result is obtained after removing the QAM symbols embedded in the proposed waveform. At this point, only three of the five pre-set point targets (A, C, and E) in the scene are clearly visible in the SAR image, while the other two targets appear defocused. Additionally, there is still a ghosting issue, as an interference strip identical to the real target appears in the range dimension.

Fig.\ref{fig7}(c) shows the SAR image obtained after compensating the FIM-LFM signal for Doppler shifts using only Algorithm 1. Although all targets remain defocused, the number of strips is reduced to one. Finally, by first removing the QAM symbols from the waveform, then using Algorithm 1 to compensate for the Doppler shifts of different sub-pulses, and finally applying the RD algorithm, a well-focused SAR image is obtained, as shown in Fig.\ref{fig7}(d). It can be observed that all five pre-set point targets in the scene are clearly displayed in the SAR image.

\begin{figure*}[htp]
	\centering
	\subfigure[Target A]{
		{\includegraphics[width=0.31\textwidth]{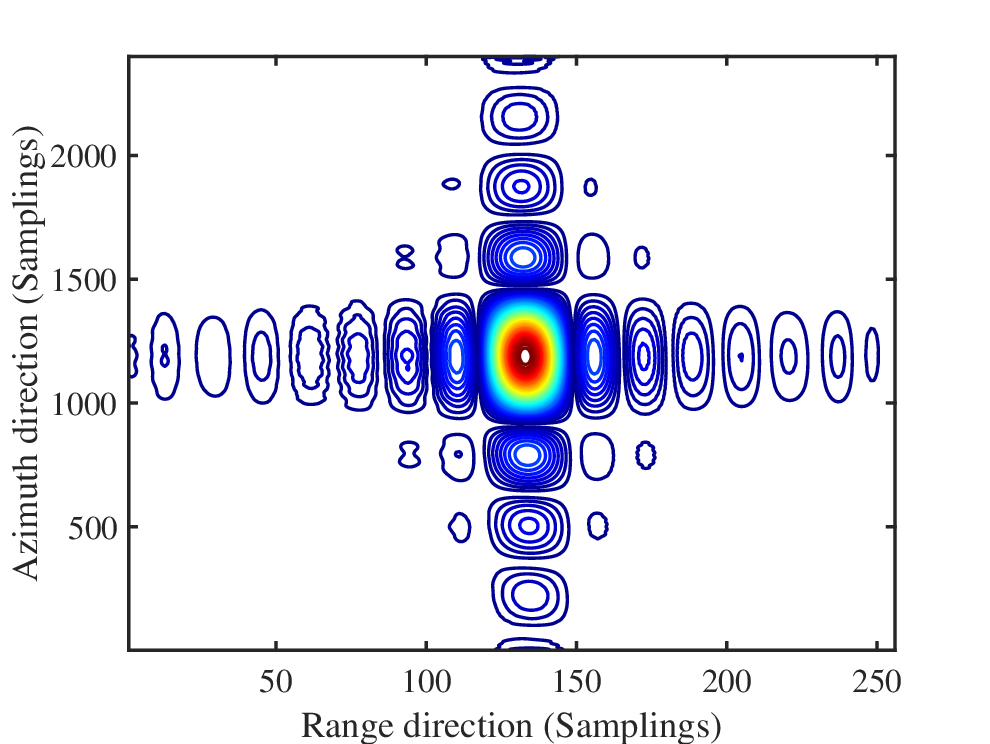}}}
	\subfigure[Target B]{
		{\includegraphics[width=0.31\textwidth]{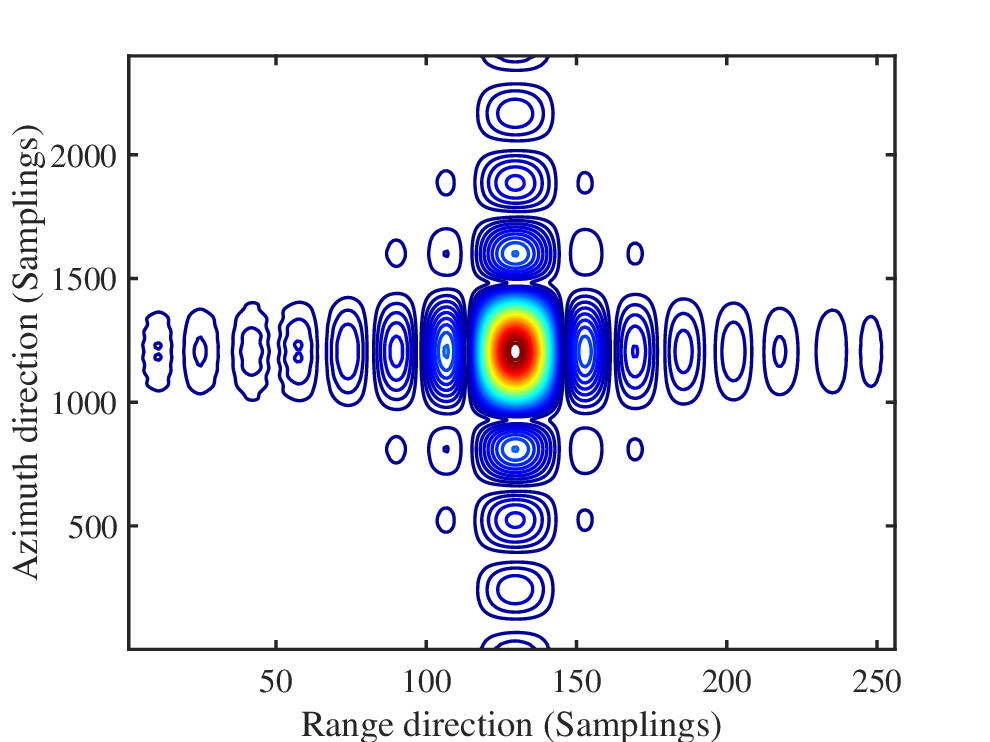}}}
	\subfigure[Target C]{
		{\includegraphics[width=0.31\textwidth]{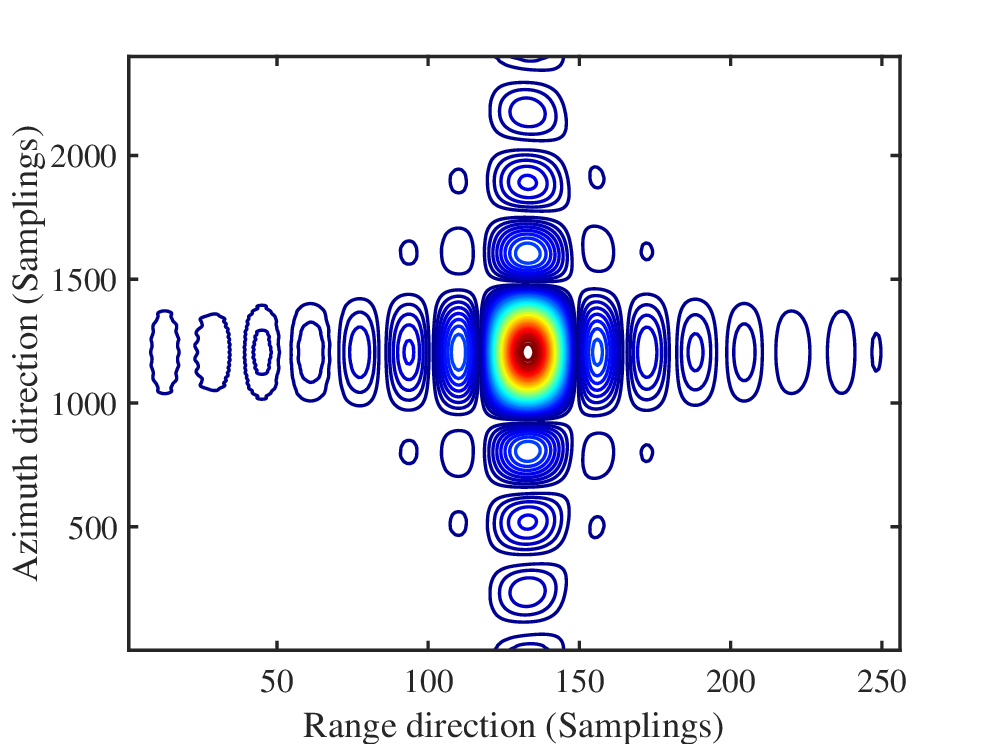}}}
%	\subfigure[After removing QAM symbol and applying compensation algorithm 1]{
%		{\includegraphics[width=0.31\textwidth]{results/results_method.eps}}}	
	\caption{Magnified image of target for Fig.\ref{fig6}.}
	\label{fig8}
\end{figure*}
\begin{figure*}[htp]
	\centering
	\subfigure[Target A]{
		{\includegraphics[width=0.31\textwidth]{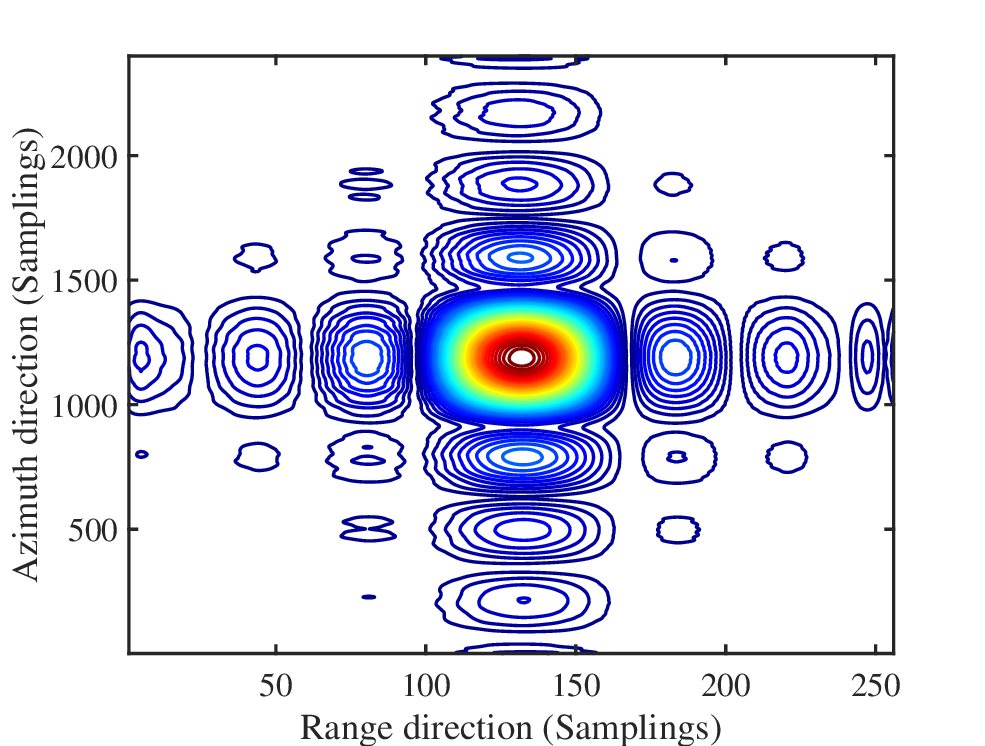}}}
	\subfigure[Target B]{
		{\includegraphics[width=0.31\textwidth]{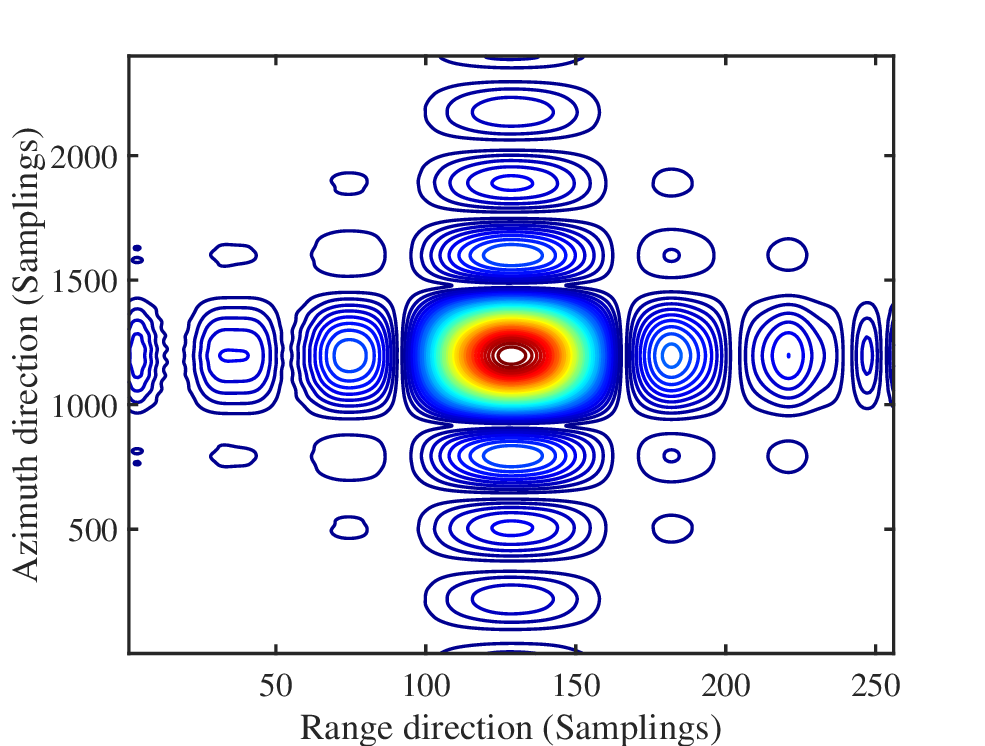}}}
	\subfigure[Target C]{
		{\includegraphics[width=0.31\textwidth]{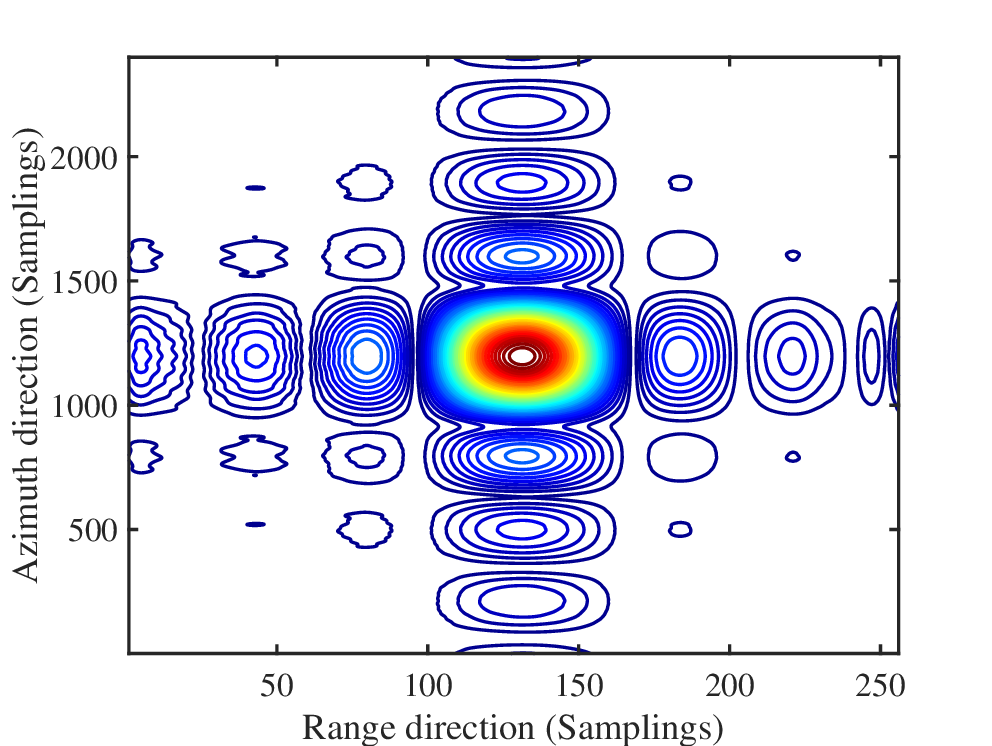}}}
	%	\subfigure[After removing QAM symbol and applying compensation algorithm 1]{
		%		{\includegraphics[width=0.31\textwidth]{results/results_method.eps}}}	
	\caption{Magnified image of target for Fig.\ref{fig7}.}
	\label{fig9}
\end{figure*}

To further evaluate the advantages and disadvantages of the LFM waveform and the FIM-LFM waveform proposed in this paper in terms of SAR imaging performance, Figures \ref{fig8} and \ref{fig9} present magnified images of three randomly selected points (A, B, C) from Figures \ref{fig6} and \ref{fig7}, respectively. Through comparative analysis of Figures \ref{fig8} and \ref{fig9}, it can be observed that the main lobe in the range dimension of the SAR image generated using the FIM-LFM waveform proposed in this paper has broadened, indicating a decrease in resolution compared to the SAR targets obtained with the LFM waveform. However, the change in the main lobe in the azimuth dimension is minimal, suggesting that the resolution difference between the two waveforms in the azimuth dimension is not significant.

\begin{figure*}[htp]
	\centering
	\subfigure[Target A]{
		{\includegraphics[width=0.31\textwidth]{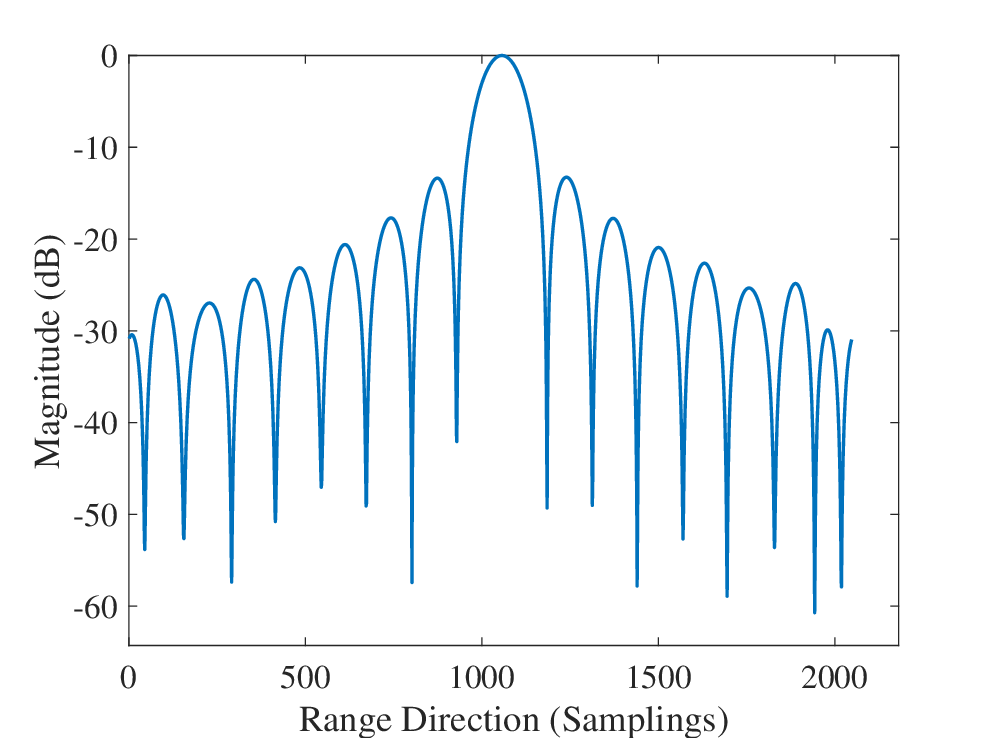}}}
	\subfigure[Target B]{
		{\includegraphics[width=0.31\textwidth]{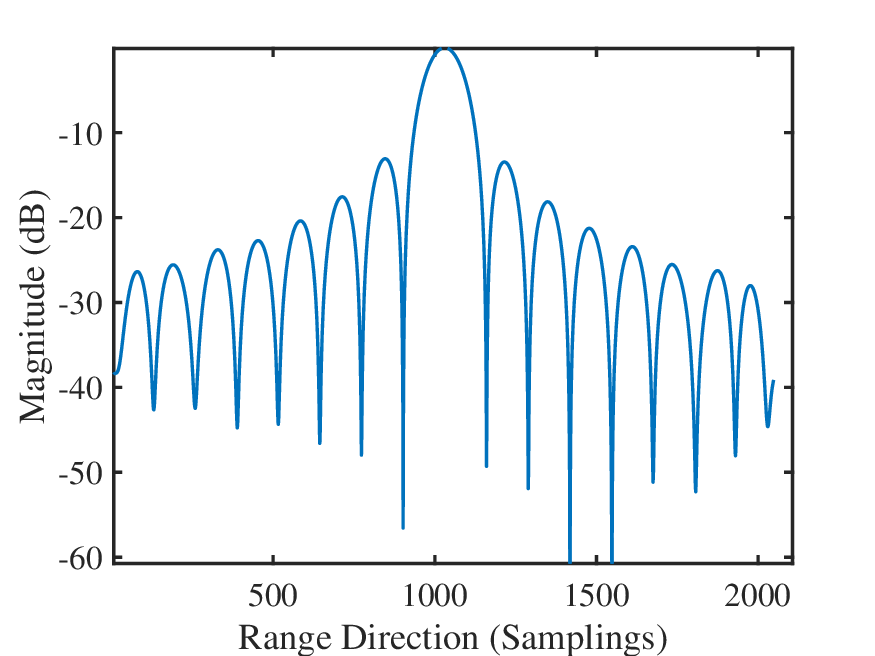}}}
	\subfigure[Target C]{
		{\includegraphics[width=0.31\textwidth]{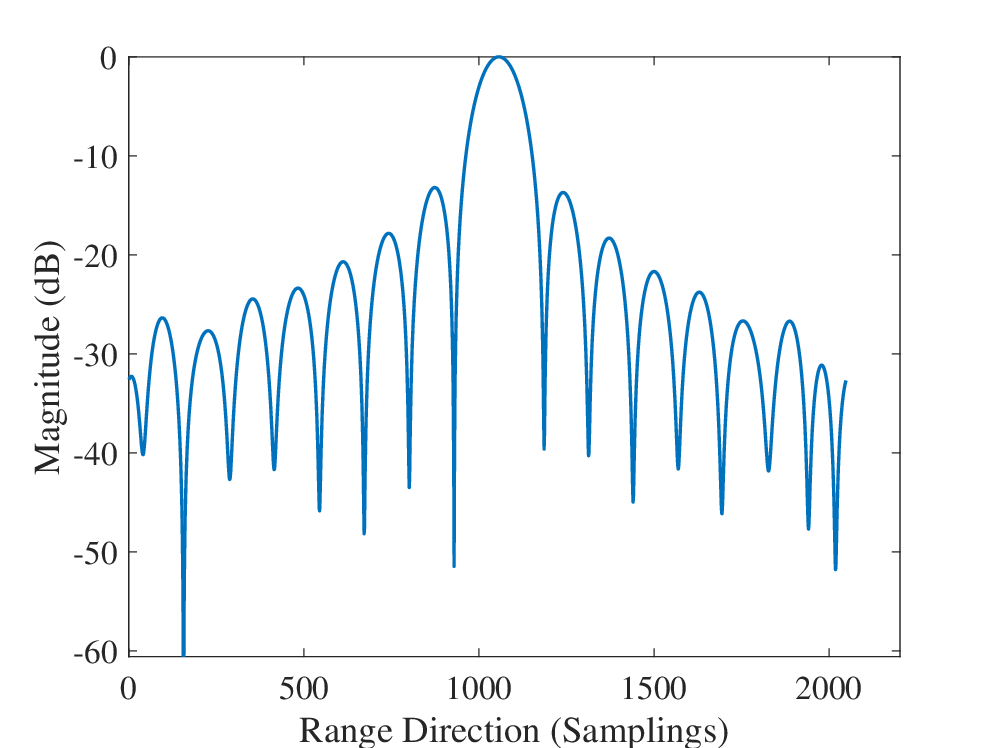}}}
	%	\subfigure[After removing QAM symbol and applying compensation algorithm 1]{
		%		{\includegraphics[width=0.31\textwidth]{results/results_method.eps}}}	
	\caption{Range profile for Fig.\ref{fig8}.}
	\label{fig12}
\end{figure*}

\begin{figure*}[htp]
	\centering
	\subfigure[Target A]{
		{\includegraphics[width=0.31\textwidth]{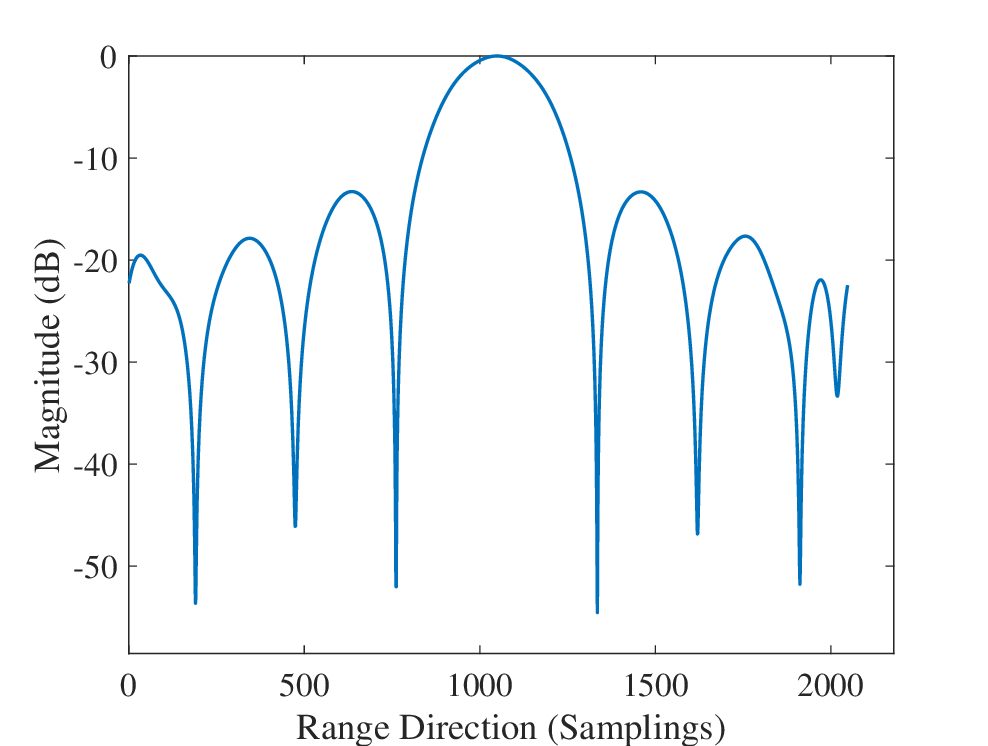}}}
	\subfigure[Target B]{
		{\includegraphics[width=0.31\textwidth]{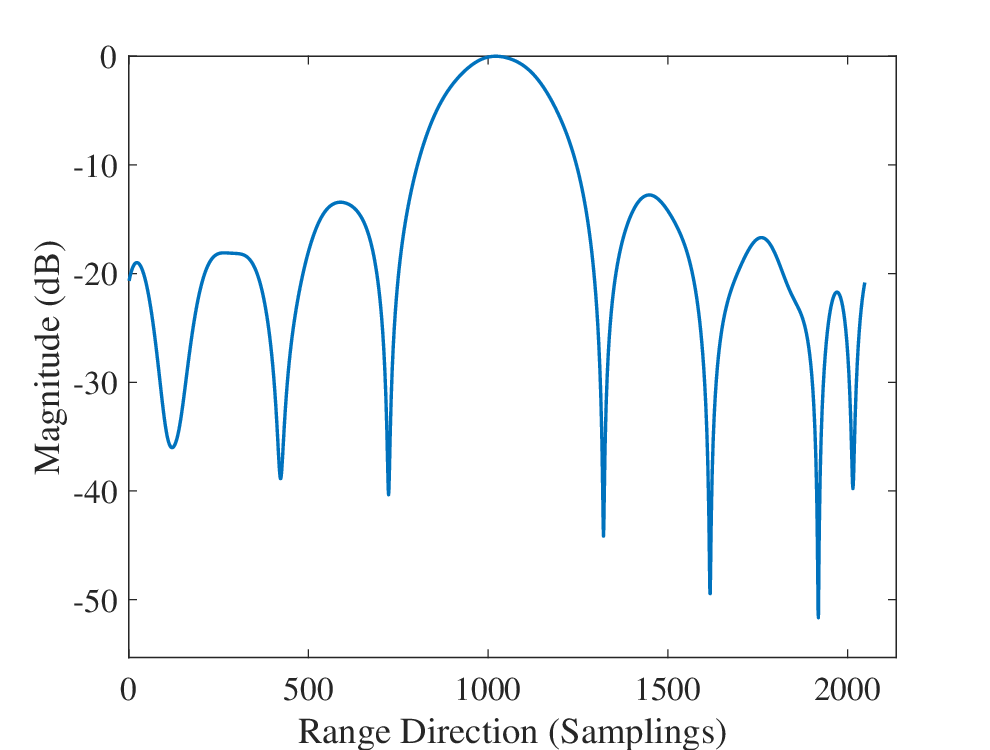}}}
	\subfigure[Target C]{
		{\includegraphics[width=0.31\textwidth]{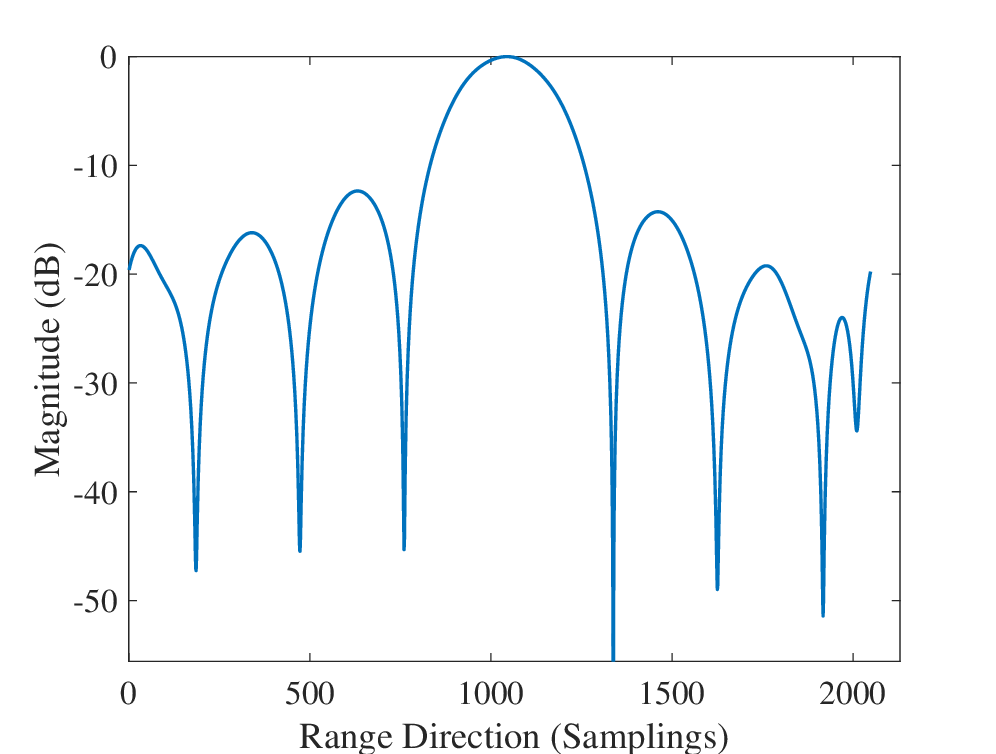}}}
	%	\subfigure[After removing QAM symbol and applying compensation algorithm 1]{
		%		{\includegraphics[width=0.31\textwidth]{results/results_method.eps}}}	
	\caption{Range profile for Fig.\ref{fig9}.}
	\label{fig13}
\end{figure*}

\begin{figure*}[htp]
	\centering
	\subfigure[Target A]{
		{\includegraphics[width=0.31\textwidth]{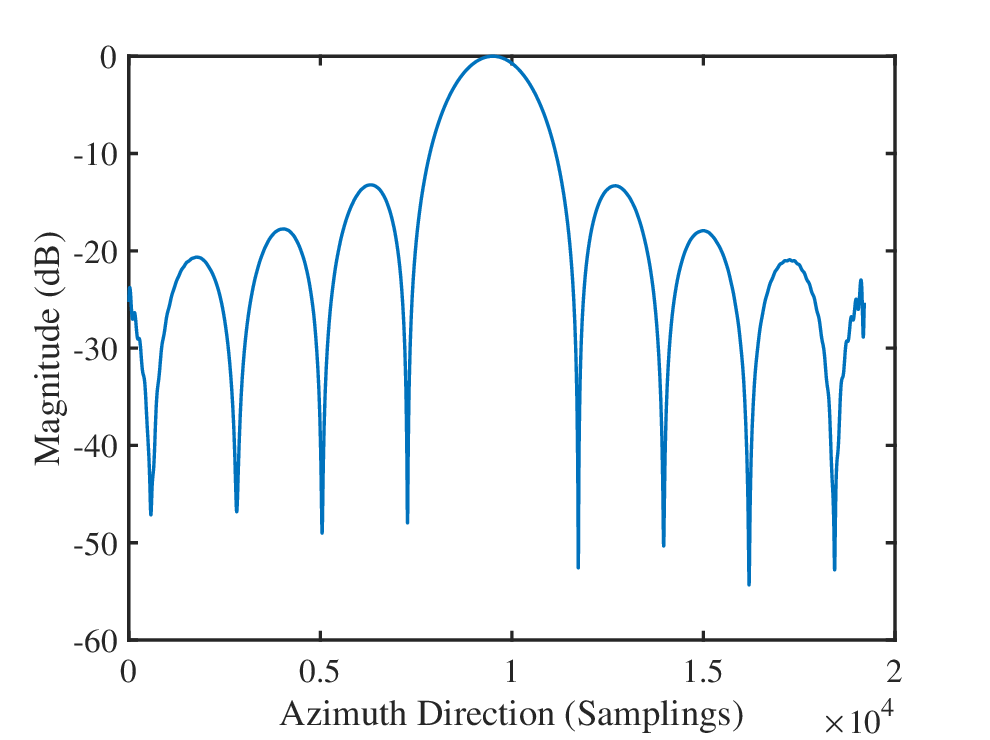}}}
	\subfigure[Target B]{
		{\includegraphics[width=0.31\textwidth]{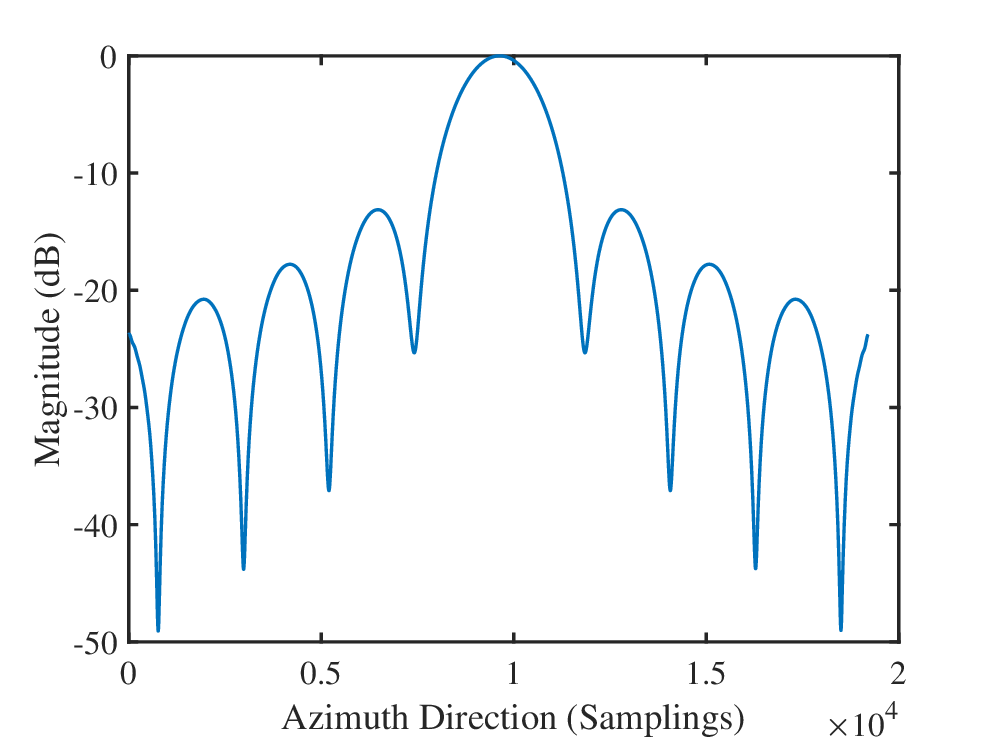}}}
	\subfigure[Target C]{
		{\includegraphics[width=0.31\textwidth]{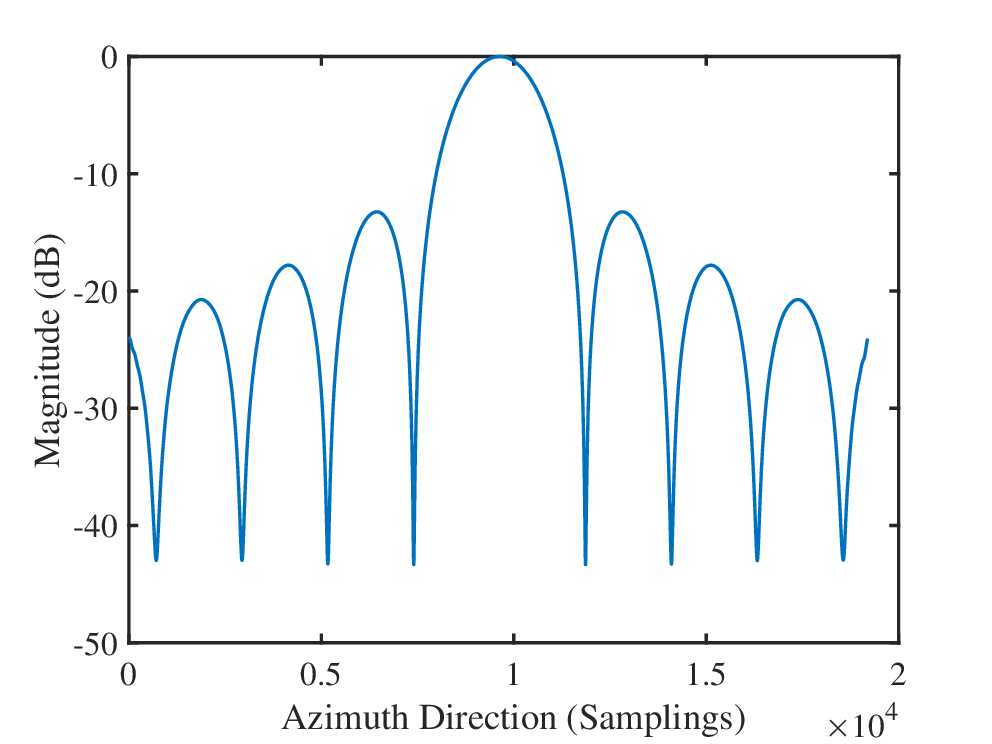}}}
	%	\subfigure[After removing QAM symbol and applying compensation algorithm 1]{
		%		{\includegraphics[width=0.31\textwidth]{results/results_method.eps}}}	
	\caption{Azimuth profile corresponding to Fig.\ref{fig8}.}
	\label{fig10}
\end{figure*}

\begin{figure*}[htp]
	\centering
	\subfigure[Target A]{
		{\includegraphics[width=0.31\textwidth]{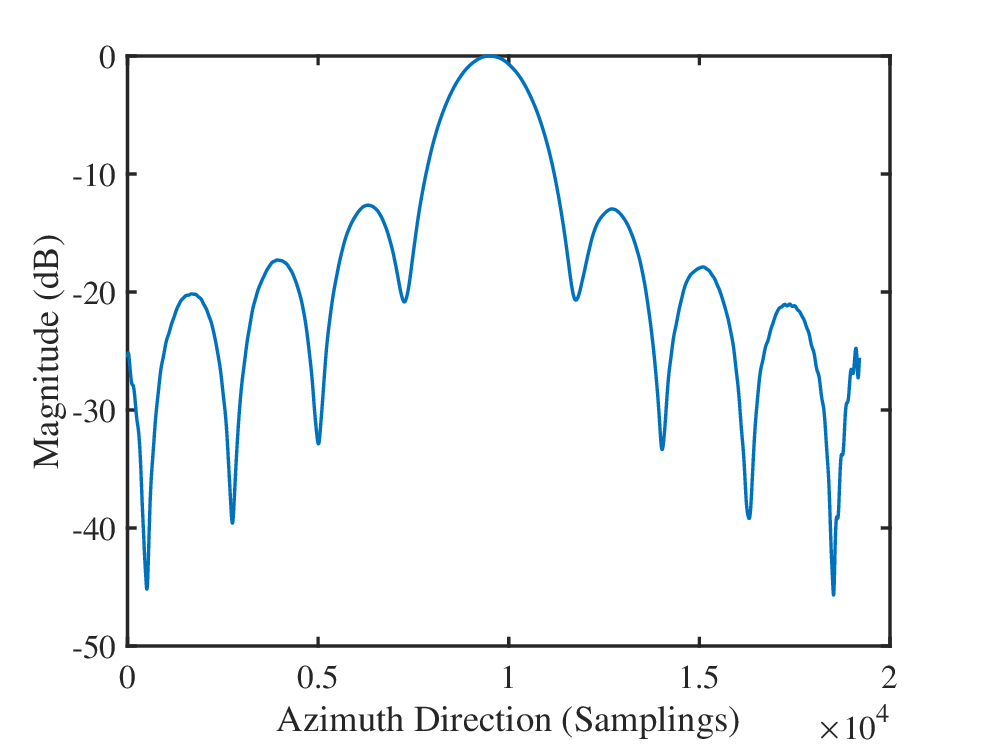}}}
	\subfigure[Target B]{
		{\includegraphics[width=0.31\textwidth]{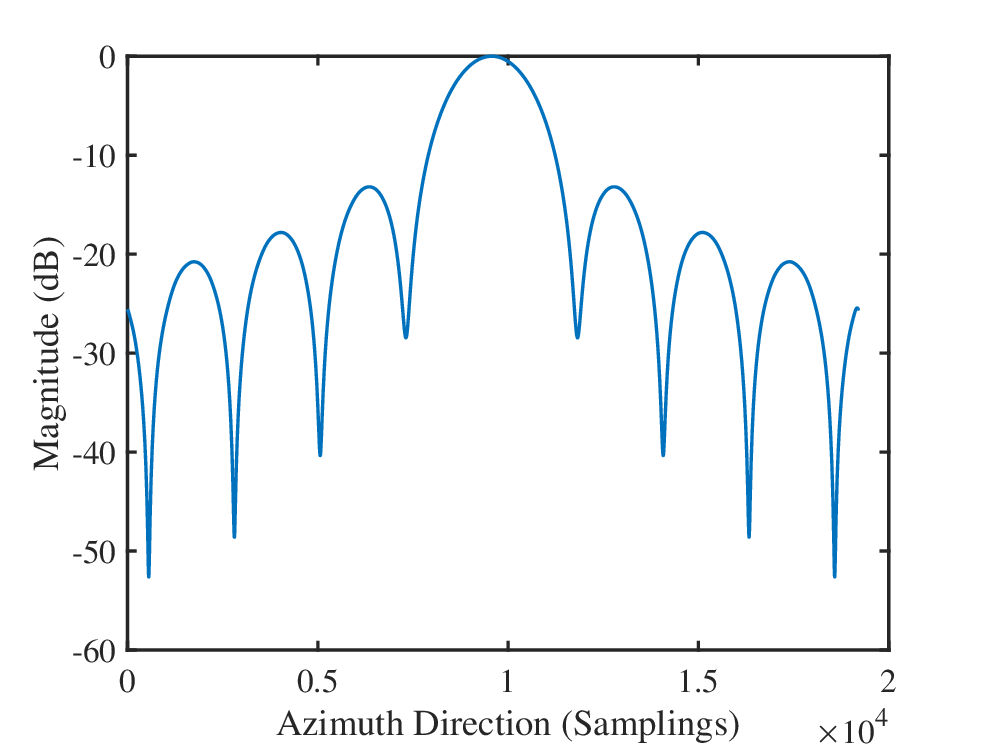}}}
	\subfigure[Target C]{
		{\includegraphics[width=0.31\textwidth]{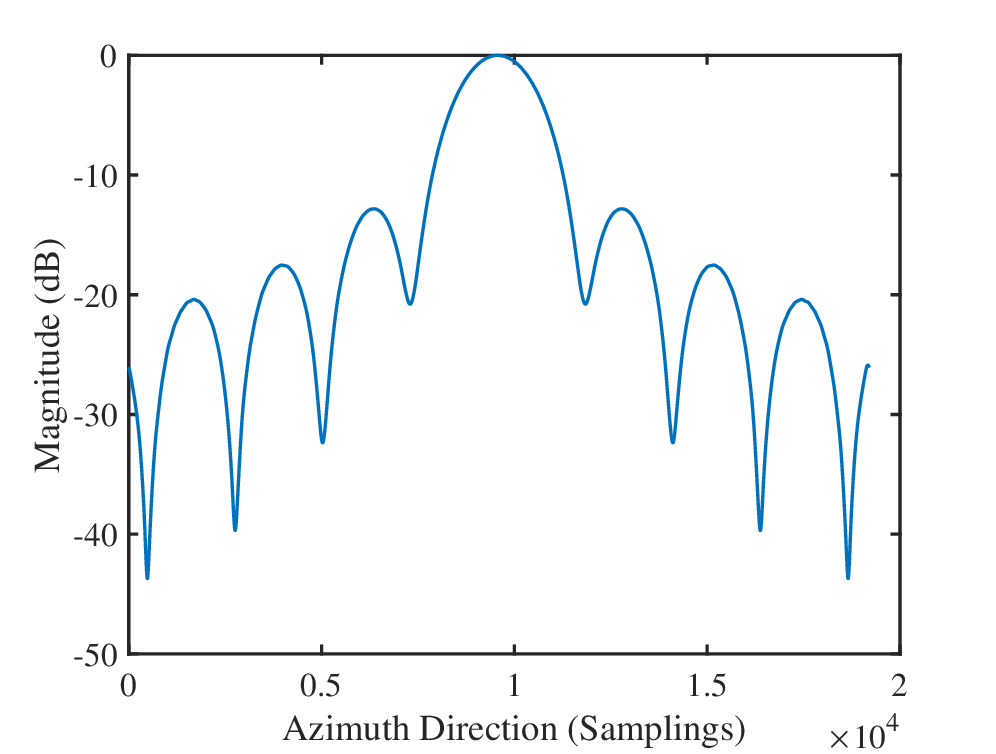}}}
	%	\subfigure[After removing QAM symbol and applying compensation algorithm 1]{
		%		{\includegraphics[width=0.31\textwidth]{results/results_method.eps}}}	
	\caption{Azimuth profile corresponding to Fig.\ref{fig9}.}
	\label{fig11}
\end{figure*}

To further evaluate the advantages and disadvantages of the LFM waveform and the FIM-LFM waveform proposed in this paper in terms of SAR imaging performance, Figures \ref{fig8} and \ref{fig9} present magnified images of three randomly selected points (A, B, C) from Figures \ref{fig6} and \ref{fig7}, respectively. Through comparative analysis of Figures \ref{fig8} and \ref{fig9}, it can be observed that the main lobe in the range dimension of the SAR image generated using the FIM-LFM waveform has broadened, indicating a decrease in resolution compared to the SAR targets obtained with the LFM waveform. This is also reflected in the range profiles shown in Figures \ref{fig12} and \ref{fig13}. However, the change in the main lobe in the azimuth dimension is minimal, suggesting that the resolution difference between the two waveforms in the azimuth dimension is not significant. Nonetheless, the azimuth profiles in Figures \ref{fig10} and \ref{fig11} indicate variations in the sidelobe depth for targets A and C, which may impact the quality of SAR imaging.

While the above conclusions provide a qualitative analysis, quantitative analysis is introduced to compare the imaging quality differences between the proposed waveform and the LFM waveform using two common metrics in the SAR field: peak sidelobe ratio (PSLR) and integration sidelobe ratio (ISLR). PSLR is defined as the ratio of the highest sidelobe peak value to the main lobe peak value of the point target impulse response, determining the ability of strong targets to "mask" weaker targets. ISLR is defined as the ratio of sidelobe energy to main lobe energy, quantitatively describing the extent to which a locally darker area is "drowned" by energy leakage from surrounding brighter areas. A smaller ISLR value indicates higher image quality. These quantitative metrics provide a basis for further analysis of the differences in imaging quality between the two waveforms.
Table \ref{tab1} presents the resolution, PSLR, and ISLR assessment results for the three point targets in Figures 8 and 9.

From the table \ref{tab1}, it is evident that the range resolution of SAR targets using LFM waveforms closely approaches the ideal resolution, with both ISLR and PSLR metrics showing favorable results. This indicates that the imaging performance of LFM waveforms is quite good. Additionally, as discussed earlier, the SAR target resolution of the proposed method is expected to range between 1.8750 and 7.5000. The results presented in Table 1 further validate the correctness of the earlier theoretical framework. Moreover, the PSLR performance of the two A, C targets in the azimuth dimension has experienced a decline, reinforcing the accuracy of the previous qualitative analysis. It is important to note that the choice of waveform does not influence the azimuth resolution of SAR, as this is primarily determined by the antenna aperture.

The analysis of the magnified images further confirms that the FIM-LFM waveform proposed in this paper experiences some loss of resolution in the range dimension compared to the LFM waveform, but it maintains good performance in the azimuth dimension. This provides valuable insights for the practical application of the FIM-LFM waveform, especially when considering scenarios that require a balance between communication and other imaging requirements, where the advantages of the FIM-LFM waveform become more apparent.
\begin{table*}[htp]
	\centering
	\caption{Evaluation of imaging results}
	\label{tab1}
	\begin{tabular}{c|c|c|c|c|c|c}
		\hline
		\multirow{3}*{Targets} 
		&\multicolumn{6}{c}{LFM}\\%&\multicolumn{6}{c}{FH-LFM}\\
		\cline{2-7}
		&\multicolumn{3}{c|}{Range domain}&\multicolumn{3}{c}{Azimuth domain}\\ %&\multicolumn{3}{c|}{Range domain}&\multicolumn{3}{c}{Azimuth domain}
		\cline{2-7}
	    %&	Resolution(m)	&	ISLR(dB)&	PSLR(dB)&	Resolution(m) &	ISLR(dB)&	PSLR(dB) 
	    &	Resolution(m)	&	ISLR(dB)&	PSLR(dB)&	Resolution(m) &	ISLR(dB) &PSLR(dB)\\
		\cline{1-7}		
		Target A	& 1.8848 &-10.2973	&-13.2654	&3.4759&-10.9346	&-13.2654\\
%				\cline{1-7}
%		%	\cline{1-5}
		Target B		& 1.9013	&-10.2907	&-13.0704	&3.4626	&-10.7574	&-13.1293\\
		Target C		&1.8848	&-10.5886	&-13.2011	&3.4812	&-10.9065	&-13.2490
		\\
		\cline{1-7}
		\multirow{3}*{Targets} 	&\multicolumn{6}{c}{Our proposed waveform}\\%&\multicolumn{6}{c}{FH-LFM}\\
		\cline{2-7}
		&\multicolumn{3}{c|}{Range domain}&\multicolumn{3}{c}{Azimuth domain}\\ %&\multicolumn{3}{c|}{Range domain}&\multicolumn{3}{c}{Azimuth domain}
		\cline{2-7}
		%&	Resolution(m)	&	ISLR(dB)&	PSLR(dB)&	Resolution(m) &	ISLR(dB)&	PSLR(dB) 
		&	Resolution(m)	&	ISLR(dB)&	PSLR(dB)&	Resolution(m) &	ISLR(dB) &PSLR(dB)\\
		\cline{1-7}		
		Target A	&4.1994	&-11.2062	&-13.2895	&3.5324	&-10.4147	&-12.6470
		 \\
		Target B	&4.3978	&-11.0788	&-12.7648	&3.5200	&-10.8710	&-13.1981
		\\
		Target C	&4.2325	&-10.9507	&-12.3531	&3.5588	&-10.3953	&-12.8217
		\\
		\hline
	\end{tabular}
	\label{eva}
\end{table*}

\subsection{Communication end}

\begin{figure}[htp]
	\centering
	{\includegraphics[width=0.45\textwidth]{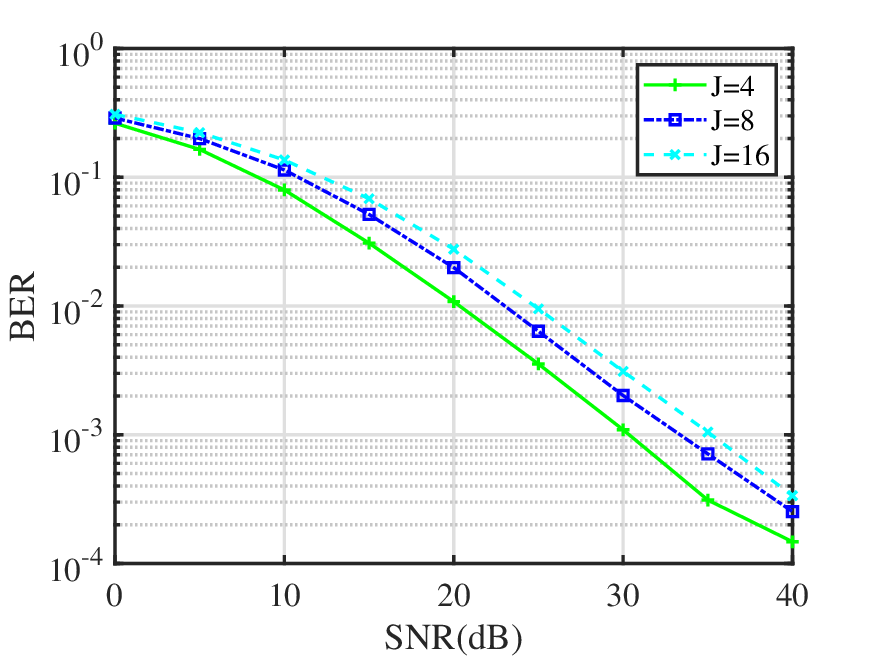}}
	\caption{BER performance by applying our system versus SNR.}
	\label{fig14}
\end{figure}

\begin{figure}[htp]
	\centering
	{\includegraphics[width=0.45\textwidth]{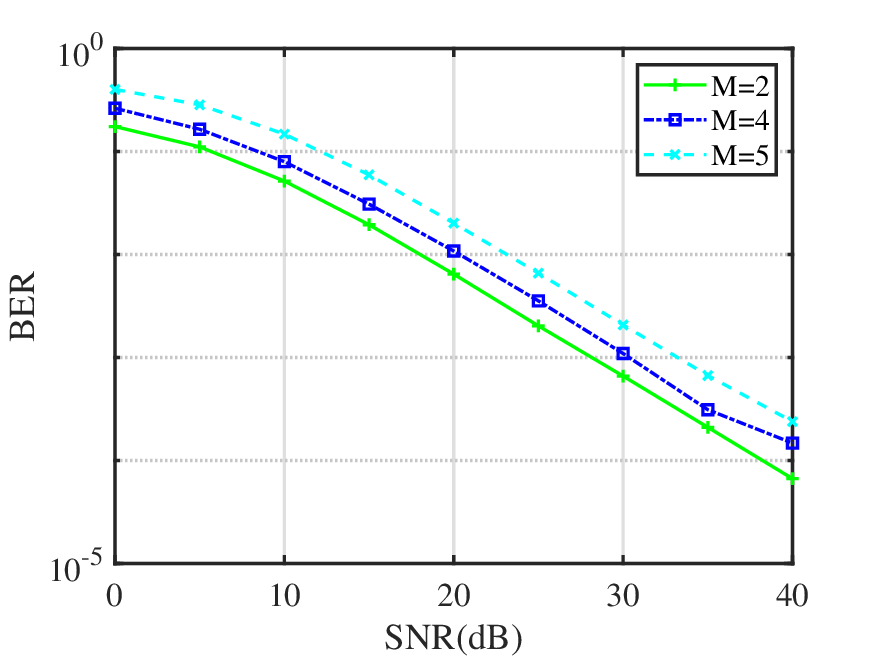}}
	\caption{BER performance by applying our system versus SNR with different sub-pulses.}
	\label{fig15}
\end{figure}

Unless stated otherwise, the parameter settings remain consistent with those established earlier. The signal-to-noise ratio (SNR) is defined in decibels (dB) as: $
\mathrm{SNR}=10\log \frac{P^2}{N_0}
$. Besides, set $\sigma ^2=1$.  Moreover, Figure \ref{fig14} presents the BER performance of the proposed ISARAC system across different QAM modulation orders as a function of SNR. The figure clearly shows that with increasing SNR, the BER decreases, reflecting an enhancement in system performance. Additionally, it is evident that as the modulation order rises, the system transmits more information, but this also leads to a corresponding increase in the error rate. Fig. \ref{fig15} illustrates the performance of the Bit Error Rate (BER) as a function of Signal-to-Noise Ratio (SNR) under various sub-pulse conditions, with the QAM order set to 4. The graph indicates that as the number of sub-pulses increases, the BER of the proposed system at the communication end tends to rise. This phenomenon can be attributed to the fact that an increase in the number of sub-pulses results in a greater number of bits being transmitted per pulse. For instance, when \( M=2 \), the number of transmitted bits is \( p=6 \), while for \( M=5 \), \( p=20 \).
\section{Conclusion}
\label{sec5}

This paper presented a novel waveform design for ISARAC on HAPs platforms, based on FIM with LFM frequency-hopping signals. By segmenting each LFM pulse into multiple sub-pulses with distinct carrier frequencies, the proposed approach achieves effective frequency hopping, enabling simultaneous SAR imaging and communication capabilities. This design not only embeds communication information within varying transmission frequencies but also reduces ADC sampling requirements in the SAR processing chain while preserving range resolution. The derivation of the ambiguity function and analysis of Doppler and range resolution provide a theoretical foundation, with established bounds for range resolution, which further supports the design's effectiveness. Additionally, a phase compensation algorithm is introduced to address phase discontinuities between sub-pulses, ensuring coherent SAR signal processing, while a two-step ML algorithm enables demodulation of QAM and FIM index symbols in the communication receiver. Numerical simulations validate the theoretical analysis and confirm the feasibility and effectiveness of the proposed waveform design, marking a significant step forward in HAPs platform radar-communication integration.

\appendices
\section{Proof of Eq.\eqref{eq15}}
\label{refA}

$\chi \left( \tau ,\xi \right)$ can be further written as
\begin{equation}
	\label{eq8}
	\begin{split}
		\chi \left( \tau ,\xi \right) =&P\int_{-\infty}^{+\infty}{\sum_{m=1}^M{p\left( \frac{t-\bigtriangleup t_m}{T_s} \right)}}
		\\
		\,\,        & 
	\times	e^{j2\pi \left( f_c+a_mB_s+\frac{1}{2}K\left( t-\bigtriangleup t_m \right) \right) \left( t-\bigtriangleup t_m \right)}\\
		&\times \sum_{m^{\prime}=1}^M{p\left( \frac{t-\bigtriangleup t_{m^{\prime}}+\tau}{T_s} \right)}\mathrm{e}^{j2\pi \xi t}
		\\
		\,\,        & \times 
		e^{-j2\pi \left( f_c+a_{m^{\prime}}B_s+\frac{1}{2}K\left( t+\tau -\bigtriangleup t_{m^{\prime}} \right) \right) \left( t+\tau -\bigtriangleup t_{m^{\prime}} \right)}
		dt
		\\
		\,\,     =&P\sum_{m=0}^{M-1}{\sum_{m^{\prime}=0}^{M-1}{e^{-j2\pi \left( f_c+a_{m^{\prime}}B_s \right) \tau}}}\check{\chi}\left( \tau ,\xi \right) 
		, 
	\end{split}
\end{equation}
where the expression of $\check{\chi}\left( \tau ,\xi \right)$ is 
\begin{equation}
	\label{eq9}
	\begin{split}
		\check{\chi}\left( \tau ,\xi \right) =&\int_{-\infty}^{+\infty}{p\left( \frac{t-\bigtriangleup t_m}{T_s} \right) p\left( \frac{t+\tau -\bigtriangleup t_{m^{\prime}}}{T_s} \right)}
		\\
		\,\,       &\times e^{j2\pi f_c\left( \bigtriangleup t_{m^{\prime}}-\bigtriangleup t_m \right)}e^{j2\pi \left( a_m-a_{m^{\prime}} \right) B_st}e^{j\pi K\left( t-\bigtriangleup t_m \right) ^2}
		\\
		\,\,          &\times e^{j2\pi \left( a_m\bigtriangleup t_m-a_{m^{\prime}}\bigtriangleup t_{m^{\prime}} \right) B_s}e^{-j\pi K\left( t-\bigtriangleup t_{m^{\prime}}+\tau \right) ^2}\mathrm{e}^{j2\pi \xi t}dt
		.
	\end{split}
\end{equation}
Besides, \eqref{eq8} indicates that $
\chi \left( \tau ,\xi \right) 
$ consists of the principal term $
\tilde{\chi}\left( \tau ,\xi \right) 
$ and the coupling term $
\bar{\chi}\left( \tau ,\xi \right) 
$, namely
\begin{equation}
	\label{key}
	\chi \left( \tau ,\xi \right) =\tilde{\chi}\left( \tau ,\xi \right) +\bar{\chi}\left( \tau ,\xi \right), 
\end{equation}
with 
\begin{align}
	\label{eq13}
	\tilde{\chi}\left( \tau ,\xi \right) =&
	\sum_{m=0}^{M-1}{e^{-j2\pi \left( f_c+a_mB_s \right) \tau}}\check{\chi}\left( \tau ,\xi \right) 
	,
	\\
	\bar{\chi}\left( \tau ,\xi \right) =&b\sum_{m=1}^M{\sum_{m^{\prime}=1,m\ne m' }^M{e^{-j2\pi \left( f_c+a_mB_s \right) \tau}\check{\chi}\left( \tau ,\xi \right)}}.	
\end{align}

The principal term, where \( m = m' \), defines the radar's resolution. In contrast, the coupling term, where \( m \neq m' \), usually serves as interference and has a lesser impact on resolution. Then we can focus our main efforts on the principal term. In this respect, 

\begin{equation}
	\label{eq44}
\begin{split}
	\check{\chi}\left( \tau ,\xi \right) =&\int_{-\infty}^{+\infty}{p\left( \frac{t-\bigtriangleup t_m}{T_s} \right) p\left( \frac{t+\tau -\bigtriangleup t_m}{T_s} \right)}
	\\
	\,\,      & \times e^{j\pi K\left( t-\bigtriangleup t_m \right) ^2}e^{-j\pi K\left( t-\bigtriangleup t_m+\tau \right) ^2}\mathrm{e}^{j2\pi \xi t}dt\\
	=&\int_{-\infty}^{+\infty}{p\left( \frac{t-\bigtriangleup t_m}{T_s} \right) p\left( \frac{t+\tau -\bigtriangleup t_m}{T_s} \right)}
	\\
	\,\,       &\times e^{-j\pi K\tau ^2}e^{-j\pi K\left( t-\bigtriangleup t_m \right) \tau}\mathrm{e}^{j2\pi \xi t}dt\\
\overset{t=t-\bigtriangleup t_m}{=}&\int_{-\infty}^{+\infty}{p\left( \frac{t}{T_s} \right) p\left( \frac{t+\tau}{T_s} \right)}
\\
\,\,       &\times e^{-j\pi K\tau ^2}e^{-j2\pi Kt\tau}\mathrm{e}^{j2\pi \xi \left( t-\bigtriangleup t_m \right)}dt.
\end{split}
\end{equation}
The expression \eqref{eq44} indicates that different values of \(\tau\) will affect the limits of integration. To ensure the existence of the integral, the condition $
\left| \tau \right|<T_s
$ must be satisfied. Therefore, we should explore expression (44) based on the different values of \(\tau\):
\subsubsection{Case 1}When $
0\leqslant 
\tau <T_s
$, \eqref{eq44} can be further expressed as 
\begin{equation}
	\label{eq45}
	\begin{split}
		\check{\chi}\left( \tau ,\xi \right) =&e^{-j\pi K\tau ^2}\mathrm{e}^{-j2\pi \xi \bigtriangleup t_m}\int_{-\frac{1}{2}T_s}^{\frac{1}{2}T_s-\tau}{e^{j2\pi t\left( \xi -K\tau \right)}}dt
		\\
		\,\,   =&\left( T_s-\tau \right) \sin c\left[ \pi \left( \xi -K\tau \right) \left( T_s-\tau \right) \right] 
		\\
		\,\,        &\times e^{-j\pi K\tau ^2}\mathrm{e}^{-j2\pi \xi \bigtriangleup t_m}e^{-j\pi \left( \xi -K\tau \right) \tau}.
	\end{split}
\end{equation}

Moreover, inserting \eqref{eq45} into \eqref{eq13} gives 
\begin{equation}
	\label{key}
	\begin{split}
		\tilde{\chi}\left( \tau ,\xi \right) =&\left( T_s-\tau \right) \sin c\left[ \pi \left( \xi -K\tau \right) \left( T_s-\tau \right) \right] 
		\\
		\,\,        &\times \sum_{m=0}^{M-1}{e^{-j2\pi \left( f_c+a_mB_s \right) \tau}}\mathrm{e}^{j2\pi \xi mT_s}e^{-j\pi \xi \tau}
	\end{split}
\end{equation}

\subsubsection{Case 2} If $
-T_s<\tau <0
$, we can rewrite \eqref{eq44} as 
\begin{equation}
	\label{eq47}
	\begin{split}
		\check{\chi}\left( \tau ,\xi \right) =&e^{-j\pi K\tau ^2}\mathrm{e}^{-j2\pi \xi \bigtriangleup t_m}\int_{-\frac{1}{2}T_s-\tau}^{\frac{1}{2}T_s}{e^{j2\pi t\left( \xi -K\tau \right)}}dt
		\\
		\,\,   =&\left( T_s+\tau \right) \sin c\left[ \pi \left( \xi -K\tau \right) \left( T_s+\tau \right) \right] 
		\\
		\,\,        &\times e^{-j\pi K\tau ^2}\mathrm{e}^{-j2\pi \xi \bigtriangleup t_m}e^{-j\pi \left( \xi -K\tau \right) \tau}.
	\end{split}
\end{equation}

Further, applying \eqref{eq47} and \eqref{eq13} results in 
\begin{equation}
	\label{key}
	\begin{split}
		\tilde{\chi}\left( \tau ,\xi \right) =&\left( T_s+\tau \right) \sin c\left[ \pi \left( \xi -K\tau \right) \left( T_s+\tau \right) \right] 
		\\
		\,\,        &\times \sum_{m=0}^{M-1}{e^{-j2\pi \left( f_c+a_mB_s \right) \tau}}\mathrm{e}^{j2\pi \xi mT_s}e^{-j\pi \xi \tau}.
	\end{split}
\end{equation}

Next, by combining all the above results, we can obtain expression \eqref{eq15}.
%\section{Proof for GCIM-MASM System}
%\label{refB}
\bibliographystyle{IEEEtran}
\bibliography{ref}

\begin{thebibliography}{10}
\providecommand{\url}[1]{#1}
\csname url@samestyle\endcsname
\providecommand{\newblock}{\relax}
\providecommand{\bibinfo}[2]{#2}
\providecommand{\BIBentrySTDinterwordspacing}{\spaceskip=0pt\relax}
\providecommand{\BIBentryALTinterwordstretchfactor}{4}
\providecommand{\BIBentryALTinterwordspacing}{\spaceskip=\fontdimen2\font plus
\BIBentryALTinterwordstretchfactor\fontdimen3\font minus
  \fontdimen4\font\relax}
\providecommand{\BIBforeignlanguage}[2]{{%
\expandafter\ifx\csname l@#1\endcsname\relax
\typeout{** WARNING: IEEEtran.bst: No hyphenation pattern has been}%
\typeout{** loaded for the language `#1'. Using the pattern for}%
\typeout{** the default language instead.}%
\else
\language=\csname l@#1\endcsname
\fi
#2}}
\providecommand{\BIBdecl}{\relax}
\BIBdecl

\bibitem{belmekkiCellularNetworkSky2024}
B.~E.~Y. Belmekki, A.~J. Aljohani, S.~A. Althubaity, A.~A. Harthi, K.~Bean,
  A.~Aijaz, and M.-S. Alouini, ``Cellular {{Network From}} the {{Sky}}:
  {{Toward People-Centered Smart Communities}},'' \emph{IEEE Open Journal of
  the Communications Society}, vol.~5, pp. 1916--1936.

\bibitem{kurtVisionFrameworkHigh2021}
\BIBentryALTinterwordspacing
G.~Kurt, M.~G. Khoshkholgh, S.~Alfattani, A.~Ibrahim, T.~S.~J. Darwish, M.~S.
  Alam, H.~Yanikomeroglu, and A.~Yongacoglu. A {{Vision}} and {{Framework}} for
  the {{High Altitude Platform Station}} ({{HAPS}}) {{Networks}} of the
  {{Future}}. [Online]. Available: \url{http://arxiv.org/abs/2007.15088}
\BIBentrySTDinterwordspacing

\bibitem{SongLopez2024HighAltitudePlatformStations}
T.~Song, D.~Lopez, M.~Meo, N.~Piovesan, and D.~Renga, ``High altitude platform
  stations: the new network energy efficiency enabler in the {6G} era,'' in
  \emph{2024 IEEE Wireless Communications and Networking Conference (WCNC)},
  2024, pp. 1--6.

\bibitem{dOliveira2016highaltitude}
F.~A. d’Oliveira, F.~C. L.~d. Melo, and T.~C. Devezas, ``High-altitude
  platforms—present situation and technology trends,'' \emph{Journal of
  Aerospace Technology and Management}, vol.~8, no.~3, pp. 249--262, 2016.

\bibitem{li2012high}
J.~Li, J.~Paden, C.~Leuschen, F.~Rodriguez-Morales, R.~D. Hale, E.~J. Arnold,
  R.~Crowe, D.~Gomez-Garcia, and P.~Gogineni, ``High-altitude radar
  measurements of ice thickness over the antarctic and greenland ice sheets as
  a part of operation icebridge,'' \emph{IEEE Transactions on Geoscience and
  Remote Sensing}, vol.~51, no.~2, pp. 742--754, 2012.

\bibitem{Renga2022CanHighAltitude}
D.~Renga and M.~Meo, ``Can high altitude platform stations make {6G}
  sustainable?'' \emph{IEEE Communications Magazine}, vol.~60, no.~9, pp.
  75--80, 2022.

\bibitem{wang2014high}
W.-Q. Wang and H.~Shao, ``High altitude platform multichannel {SAR} for
  wide-area and staring imaging,'' \emph{IEEE Aerospace and Electronic Systems
  Magazine}, vol.~29, no.~5, pp. 12--17, 2014.

\bibitem{YuZhang2024JointResourceAllocations}
X.~Yu, X.~Zhang, Y.~Rui, K.~Wang, X.~Dang, and M.~Guizani, ``Joint resource
  allocations for energy consumption optimization in {HAPS-aided MEC-NOMA}
  systems,'' \emph{IEEE Journal on Selected Areas in Communications}, pp. 1--1,
  2024.

\bibitem{Ahrazoglu2024MultiHAPs}
E.~S. Ahrazoglu, I.~Altunbas, and E.~Erdogan, ``{Multi-HAPS} thz satellite
  communication: Error and capacity analyses under {I/Q} imbalance,''
  \emph{IEEE Sensors Journal}, pp. 1--1, 2024.

\bibitem{Turk2024Design}
S.~E. Turk, E.~S. Ahrazoglu, E.~Erdogan, and I.~Altunbas, ``Design of energy
  efficient {Multi-HAPS} assisted hybrid rf/fso satellite communication systems
  with optimal placement,'' \emph{IEEE Transactions on Green Communications and
  Networking}, pp. 1--1, 2024.

\bibitem{LiuMasouros2021ATutorial}
F.~Liu and C.~Masouros, ``A tutorial on joint radar and communication
  transmission for vehicular networks—part i: Background and fundamentals,''
  \emph{IEEE Communications Letters}, vol.~25, no.~2, pp. 322--326, 2021.

\bibitem{ma2020joint}
D.~Ma, N.~Shlezinger, T.~Huang, Y.~Liu, and Y.~C. Eldar, ``Joint
  radar-communication strategies for autonomous vehicles: Combining two key
  automotive technologies,'' \emph{IEEE signal processing magazine}, vol.~37,
  no.~4, pp. 85--97, 2020.

\bibitem{bicua2019multicarrier}
M.~Bic{\u{a}} and V.~Koivunen, ``Multicarrier radar-communications waveform
  design for rf convergence and coexistence,'' in \emph{ICASSP 2019-2019 IEEE
  International Conference on Acoustics, Speech and Signal Processing
  (ICASSP)}.\hskip 1em plus 0.5em minus 0.4em\relax IEEE, 2019, pp. 7780--7784.

\bibitem{LiHu2024AoIAware}
Z.~Li, F.~Hu, Q.~Li, Z.~Ling, Z.~Chang, and T.~Hämäläinen, ``Aoi-aware
  waveform design for cooperative joint radar-communications systems with
  online prediction of radar target property,'' \emph{IEEE Transactions on
  Communications}, vol.~72, no.~10, pp. 6029--6043, 2024.

\bibitem{zhang2021overview}
J.~A. Zhang, F.~Liu, C.~Masouros, R.~W. Heath, Z.~Feng, L.~Zheng, and
  A.~Petropulu, ``An overview of signal processing techniques for joint
  communication and radar sensing,'' \emph{IEEE Journal of Selected Topics in
  Signal Processing}, vol.~15, no.~6, pp. 1295--1315, 2021.

\bibitem{ChakravarthiAshoka2024ASurveyon}
A.~Chakravarthi~Mahipathi, B.~Pardhasaradhi, P.~Lingadevaru, P.~Srihari,
  J.~D’Souza, and L.~Reddy~Cenkeramaddi, ``A survey on waveform design for
  radar-communication convergence,'' \emph{IEEE Access}, vol.~12, pp.
  75\,442--75\,461, 2024.

\bibitem{Ma2024IntegratedSensing}
H.~Ma, ``Integrated sensing and communication - the {ISAC} technology,'' in
  \emph{2024 IEEE 2nd International Conference on Sensors, Electronics and
  Computer Engineering (ICSECE)}, 2024, pp. 225--229.

\bibitem{LiuLiao2017AdaptiveOFDM}
Y.~Liu, G.~Liao, J.~Xu, Z.~Yang, and Y.~Zhang, ``Adaptive {OFDM} integrated
  radar and communications waveform design based on information theory,''
  \emph{IEEE Communications Letters}, vol.~21, no.~10, pp. 2174--2177, 2017.

\bibitem{XiaoZeng2022WaveformDesign}
Z.~Xiao and Y.~Zeng, ``Waveform design and performance analysis for full-duplex
  integrated sensing and communication,'' \emph{IEEE Journal on Selected Areas
  in Communications}, vol.~40, no.~6, pp. 1823--1837, 2022.

\bibitem{gaudio2019performance}
L.~Gaudio, M.~Kobayashi, B.~Bissinger, and G.~Caire, ``Performance analysis of
  joint radar and communication using ofdm and otfs,'' in \emph{2019 IEEE
  International Conference on Communications Workshops (ICC Workshops)}.\hskip
  1em plus 0.5em minus 0.4em\relax IEEE, 2019, pp. 1--6.

\bibitem{xiao2024novel}
Z.~Xiao, R.~Liu, M.~Li, Q.~Liu, and A.~L. Swindlehurst, ``A novel joint
  angle-range-velocity estimation method for {MIMO-OFDM ISAC} systems,''
  \emph{IEEE Transactions on Signal Processing}, 2024.

\bibitem{sturm2009ofdm}
C.~Sturm, T.~Zwick, and W.~Wiesbeck, ``An ofdm system concept for joint radar
  and communications operations,'' in \emph{VTC Spring 2009-IEEE 69th Vehicular
  Technology Conference}.\hskip 1em plus 0.5em minus 0.4em\relax IEEE, 2009,
  pp. 1--5.

\bibitem{hsu2021analysis}
H.-W. Hsu, M.-C. Lee, M.-X. Gu, Y.-C. Lin, and T.-S. Lee, ``Analysis and design
  for pilot power allocation and placement in {OFDM} based integrated radar and
  communication in automobile systems,'' \emph{IEEE Transactions on Vehicular
  Technology}, vol.~71, no.~2, pp. 1519--1535, 2021.

\bibitem{huang2015low}
T.~Huang and T.~Zhao, ``Low pmepr ofdm radar waveform design using the
  iterative least squares algorithm,'' \emph{IEEE Signal Processing Letters},
  vol.~22, no.~11, pp. 1975--1979, 2015.

\bibitem{wu2024low}
J.~Wu, L.~Li, W.~Lin, J.~Liang, and Z.~Han, ``Low-complexity waveform design
  for papr reduction in integrated sensing and communication systems based on
  {ADMM},'' \emph{IEEE Sensors Journal}, 2024.

\bibitem{liu2024ofdm}
F.~Liu, Y.~Zhang, Y.~Xiong, S.~Li, W.~Yuan, F.~Gao, S.~Jin, and G.~Caire,
  ``{OFDM} achieves the lowest ranging sidelobe under random {ISAC}
  signaling,'' \emph{arXiv preprint arXiv:2407.06691}, 2024.

\bibitem{wei2023waveform}
Z.~Wei, J.~Piao, X.~Yuan, H.~Wu, J.~A. Zhang, Z.~Feng, L.~Wang, and P.~Zhang,
  ``Waveform design for mimo-ofdm integrated sensing and communication system:
  An information theoretical approach,'' \emph{IEEE Transactions on
  Communications}, 2023.

\bibitem{li2024mimo}
P.~Li, M.~Li, R.~Liu, Q.~Liu, and A.~L. Swindlehurst, ``{MIMO-OFDM ISAC}
  waveform design for range-doppler sidelobe suppression,'' \emph{arXiv
  preprint arXiv:2406.17218}, 2024.

\bibitem{boroujeni2024enhancing}
A.~K. Boroujeni, G.~Bagheri, and S.~K{\"o}psell, ``Enhancing frequency hopping
  security in isac systems: A physical layer security approach,'' in \emph{2024
  IEEE 4th International Symposium on Joint Communications \& Sensing
  (JC\&S)}.\hskip 1em plus 0.5em minus 0.4em\relax IEEE, 2024, pp. 1--6.

\bibitem{wu2021frequency}
K.~Wu, J.~A. Zhang, X.~Huang, and Y.~J. Guo, ``Frequency-hopping {MIMO}
  radar-based communications: An overview,'' \emph{IEEE Aerospace and
  Electronic Systems Magazine}, vol.~37, no.~4, pp. 42--54, 2021.

\bibitem{wang2019co}
X.~Wang and J.~Xu, ``Co-design of joint radar and communications systems
  utilizing frequency hopping code diversity,'' in \emph{2019 IEEE Radar
  Conference (RadarConf)}.\hskip 1em plus 0.5em minus 0.4em\relax IEEE, 2019,
  pp. 1--6.

\bibitem{hassanien2017dual}
A.~Hassanien, B.~Himed, and B.~D. Rigling, ``A dual-function {MIMO}
  radar-communications system using frequency-hopping waveforms,'' in
  \emph{2017 IEEE Radar Conference (RadarConf)}.\hskip 1em plus 0.5em minus
  0.4em\relax IEEE, 2017, pp. 1721--1725.

\bibitem{yuan2022orthogonal}
W.~Yuan, Z.~Wei, S.~Li, R.~Schober, and G.~Caire, ``Orthogonal time frequency
  space modulation—part iii: {ISAC} and potential applications,'' \emph{IEEE
  Communications Letters}, vol.~27, no.~1, pp. 14--18, 2022.

\bibitem{zhu2023joint}
Q.~Zhu, M.~Li, R.~Liu, and Q.~Liu, ``Joint transceiver beamforming and
  reflecting design for active {RIS-aided ISAC} systems,'' \emph{IEEE
  Transactions on Vehicular Technology}, vol.~72, no.~7, pp. 9636--9640, 2023.

\bibitem{ma2021frac}
D.~Ma, N.~Shlezinger, T.~Huang, Y.~Liu, and Y.~C. Eldar, ``{FRaC}: {FMCW}-based
  joint radar-communications system via index modulation,'' \emph{IEEE journal
  of selected topics in signal processing}, vol.~15, no.~6, pp. 1348--1364,
  2021.

\bibitem{huang2020majorcom}
T.~Huang, N.~Shlezinger, X.~Xu, Y.~Liu, and Y.~C. Eldar, ``{MAJoRCom}: A
  dual-function radar communication system using index modulation,'' \emph{IEEE
  Transactions on signal processing}, vol.~68, pp. 3423--3438, 2020.

\bibitem{barrenechea2007fmcw}
P.~Barrenechea, F.~Elferink, and J.~Janssen, ``{FMCW} radar with broadband
  communication capability,'' in \emph{2007 European Radar Conference}.\hskip
  1em plus 0.5em minus 0.4em\relax IEEE, 2007, pp. 130--133.

\bibitem{gu2022design}
M.-X. Gu, M.-C. Lee, Y.-S. Liu, and T.-S. Lee, ``Design and analysis of
  frequency hopping-aided {FMCW}-based integrated radar and communication
  systems,'' \emph{IEEE Transactions on Communications}, vol.~70, no.~12, pp.
  8416--8432, 2022.

\bibitem{xie2021waveform}
R.~Xie, K.~Luo, and T.~Jiang, ``Waveform design for {LFM-MPSK-based} integrated
  radar and communication toward {IoT} applications,'' \emph{IEEE Internet of
  Things Journal}, vol.~9, no.~7, pp. 5128--5141, 2021.

\bibitem{ma2021spatial}
D.~Ma, N.~Shlezinger, T.~Huang, Y.~Shavit, M.~Namer, Y.~Liu, and Y.~C. Eldar,
  ``Spatial modulation for joint radar-communications systems: Design,
  analysis, and hardware prototype,'' \emph{IEEE Transactions on Vehicular
  Technology}, vol.~70, no.~3, pp. 2283--2298, 2021.

\bibitem{liu2023integrated}
Z.~Liu, F.~Zesong, P.~Liu, X.~Wang, Z.~Zheng, D.~Zhou, and W.~Yuan,
  ``Integrated sensing and communication for {UAV-Borne SAR} systems,'' in
  \emph{2023 22nd International Symposium on Communications and Information
  Technologies (ISCIT)}.\hskip 1em plus 0.5em minus 0.4em\relax IEEE, 2023, pp.
  1--6.

\bibitem{manzoni2024evaluation}
M.~Manzoni, S.~Moro, F.~Linsalata, M.~G. Polisano, A.~V. Monti-Guarnieri, and
  S.~Tebaldini, ``Evaluation of {UAV-Based ISAC SAR} imaging: Methods and
  performances,'' in \emph{2024 IEEE Radar Conference (RadarConf24)}.\hskip 1em
  plus 0.5em minus 0.4em\relax IEEE, 2024, pp. 1--6.

\bibitem{lahmeri2022trajectory}
M.-A. Lahmeri, W.~Ghanem, C.~Knill, and R.~Schober, ``Trajectory and resource
  optimization for {UAV} synthetic aperture radar,'' in \emph{2022 IEEE
  Globecom Workshops (GC Wkshps)}.\hskip 1em plus 0.5em minus 0.4em\relax IEEE,
  2022, pp. 897--903.

\bibitem{lahmeri2024uav}
M.-A. Lahmeri, V.~Mustieles-P{\'e}rez, M.~Vossiek, G.~Krieger, and R.~Schober,
  ``{UAV} formation and resource allocation optimization for
  communication-assisted {3D InSAR} sensing,'' \emph{arXiv preprint
  arXiv:2407.06607}, 2024.

\bibitem{zhang2014ofdm}
T.~Zhang and X.-G. Xia, ``{OFDM} synthetic aperture radar imaging with
  sufficient cyclic prefix,'' \emph{IEEE Transactions on Geoscience and Remote
  Sensing}, vol.~53, no.~1, pp. 394--404, 2014.

\bibitem{wang2019first}
J.~Wang, X.-D. Liang, L.-Y. Chen, L.-N. Wang, and K.~Li, ``First demonstration
  of joint wireless communication and high-resolution {SAR} imaging using
  airborne {MIMO} radar system,'' \emph{IEEE Transactions on Geoscience and
  Remote Sensing}, vol.~57, no.~9, pp. 6619--6632, 2019.

\bibitem{liu2022radar}
G.~Liu, Y.~Wang, and W.~Yang, ``Radar sensor and data communication system
  based on {OFDM} without cyclic prefix,'' \emph{IEEE Sensors Journal},
  vol.~23, no.~7, pp. 7578--7590, 2022.

\bibitem{Yang2023WaveformDesign}
J.~Yang, Y.~Tan, X.~Yu, G.~Cui, and D.~Zhang, ``Waveform design for watermark
  framework based {DFRC} system with application on joint {SAR} imaging and
  communication,'' \emph{IEEE Transactions on Geoscience and Remote Sensing},
  vol.~61, pp. 1--14, 2023.

\bibitem{JirousekPeichl2024DLR}
M.~Jirousek, M.~Peichl, S.~Anger, S.~Dill, and M.~Limbach, ``The {DLR} high
  altitude platform synthetic aperture radar instrument {HAPSAR},'' in
  \emph{EUSAR 2024; 15th European Conference on Synthetic Aperture Radar},
  2024, pp. 1244--1248.

\bibitem{JirousekPeichl2022SyntheticApertureRadar}
M.~Jirousek, M.~Peichl, S.~Anger, M.~Engel, S.~Dill, R.~Scheiber, and
  S.~Baumgartner, ``Synthetic aperture radar design for a high-altitude
  platform,'' in \emph{EUSAR 2022; 14th European Conference on Synthetic
  Aperture Radar}, 2022, pp. 1--4.

\bibitem{JirousekPeichl2023DesignofaSynthetic}
M.~Jirousek, M.~Peichl, S.~Anger, S.~Dill, and M.~Engel, ``Design of a
  synthetic aperture radar instrument for a high-altitude platform,'' in
  \emph{IGARSS 2023 - 2023 IEEE International Geoscience and Remote Sensing
  Symposium}, 2023, pp. 2045--2048.

\end{thebibliography}

% \newpage
% \color{red}
% \begin{eqnarray}
%      \tilde{s}_{k,m}(t,\tau(s))\left[s_{\mathrm{rf}}\left( t \right) \right]^* &=& p(t-\bigtriangleup t_{kM+m}-\tau(s))h_{kM+m}c_{kM+m} \sqrt{P}e^{j2\pi \left( f_c+a_{kM+m}B_s + \frac{K}{2}\left( t-\bigtriangleup t_{kM+m}-\tau(s)\right)\right) \left( t-\bigtriangleup t_{kM+m}-\tau(s)\right)}\notag\\
%     &&p(t-\bigtriangleup t_{kM+m}) e^{-j2\pi \left( f_c+\frac{1}{2}K\left( t-\bigtriangleup t_{kM+m} \right) \right) \left( t-\bigtriangleup t_{kM+m} \right)} \notag \\
%     %
%      &=& h_{kM+m}c_{kM+m}\sqrt{P}e^{j2\pi \left(f_c+a_{kM+m}B_s + \frac{K}{2}\left( t-\bigtriangleup t_{kM+m}-\tau(s)\right)\right) \left( t-\bigtriangleup t_{kM+m}-\tau \left( s \right) \right)}\notag\\
%     &&p(t-\bigtriangleup t_{kM+m}) e^{j2\pi \left( f_c+\frac{1}{2}K\left( t-\bigtriangleup t_{kM+m} \right) \right) \left( t-\bigtriangleup t_{kM+m} \right)}\notag \\
%     %
%      &=& h_{kM+m}c_{kM+m}\sqrt{P}e^{-j2\pi \left(\left(f_c+a_{kM+m}B_s\right)\tau(s) + K\left(t-\bigtriangleup t_{kM+m}\right)\tau(s) - \frac{K\tau^2(s)}{2} \right)}\notag
% \end{eqnarray}

\end{document}